\documentclass[%
 reprint,
 amsmath,amssymb,
 aps,
]{revtex4-1}

\usepackage{graphicx}% Include figure files
\usepackage{dcolumn}% Align table columns on decimal point
\usepackage{bm}% bold math
\usepackage{tabularx}
\usepackage{float}
\graphicspath{{fig/}}
\begin{document}

\preprint{APS/123-QED}

\title{An empirical characterization of community structures in complex networks using a bivariate map of quality metrics}% Force line breaks with \\

\author{Vinh-Loc Dao}
 %\altaffiliation{LUSSI Department - IMT Atlantique}\\
 \email{vinh.dao@imt-atlantique.fr}
\author{C\'{e}cile Bothorel}%
 \email{cecile.bothorel@imt-atlantique.fr}
\author{Philippe Lenca}%
 \email{philippe.lencal@imt-atlantique.fr}
\affiliation{%
IMT Atlantique, Lab-STICC - CNRS, UMR 6285, F-29238 Brest, France \\
}%

\date{\today}% It is always \today, today,

\begin{abstract}
Community detection emerges as an important task in the discovery of network mesoscopic structures. However, the concept of a \textit{``good"} community is very context-dependent and it is relatively complicated to deduce community characteristics using available community detection techniques. In reality, the existence of a gap between structural  goodness quality metrics and expected topological patterns creates a confusion in evaluating community structures. In this paper, we introduce an empirical multivariate analysis of different structural goodness properties in order to characterize several detectable community topologies. Specifically, we show that a combination of two representative structural dimensions including community transitivity and hub dominance allows to distinguish different topologies such as star-based, clique-based, string-based and grid-based structures. Additionally, these classes of topology disclose structural proximities with those of graphs created by Erd\H{o}s-R\'{e}nyi, Watts-Strogatz and Barab\'{a}si-Albert generative models. We illustrate popular community topologies identified by different detection methods on a large dataset composing many network categories and associate their structures with the most related graph generative model. Interestingly, this conjunctive representation sheds light on fundamental differences between mesoscopic structures in various network categories including: communication, information, biological, technological, social, ecological, synthetic networks and more. 

\begin{description}
\item[Keywords]
Complex Networks, Community Structure, Community Characterization,\\
Cluster Description, Graph Model.
\end{description}
\end{abstract}

\pacs{Valid PACS appear here}% PACS, the Physics and Astronomy
\maketitle

%\tableofcontents

\section{Introduction}
\label{sec:intro}
The representation of complex networks using graphs opens tremendous possibilities to discover their structural characteristics and latter allows to explain different phenomena related to the system functionality. Hence, an impressive amount of work have been conducted to study large-scale structure in many network categories such as social networks, \cite{leskovec:2008b},~\cite{yang:2015a}, biochemical networks,~\cite{eric:2003a},~\cite{albert:2005a},~\cite{zhu:2007a}, computer networks, \cite{faloutsos:1999a}, \cite{vazquez:2002a} etc. Among important features, community structure receives an immense attention
since it could help to discover network organization and thus allow to explain several mechanisms that affect network evolution in a mesoscopic level; to understand different dynamic processes happening on the network; to study network behaviors associating to functional blocks and so on.  
Consequently, a variety of approaches to discover different aspects of modular structure have been developed in the last few decades~\cite{raghavan:2007a},~\cite{girvan:2002a},~\cite{newman:2004a},~\cite{rosvall:2008a}. Several efforts have been also dedicated to evaluate and categorize these methods in a systematic way~\cite{leskovec:2010a},~\cite{fortunato:2010a},~\cite{coscia:2012a},~\cite{rosvall:2017a}.

Even it is widely accepted that communities are groups of nodes in a network where there are much more edges that connect nodes of the same group than edges that connect nodes in different groups~\cite{fortunato:2010a}, community quality evaluation has been always an controversial issue due to its ambiguous definition. That is why community detection is still particularly problem-driven, which means the perspective of what is expected as a \textit{good} community varies according to different contexts and there is no formal concept of community that is well mathematically formulated and widely accepted as the same time. For each specific problem, one may have a particular definition of a good community which reflects expected modular structures inside a network of interest. However, plausible solutions are generally difficult to be evaluated or validated without a presence of specialists who understand well the system in question and the discovery mechanisms of algorithms in disposition.

As a consequence, although several community detection methods are available in the literature, the question of which method to be adopted in order to find a specific structure of community remains a challenge to be solved~\cite{fortunato:2016a}. In this paper, we attempt to answer the question: \textit{``what do structural communities in real networks look like?"} by characterizing systematically communities discovered in many real world network categories. This characterization helps network analysts to discern detectable types of community structures that could be found on their networks. The expected structure may exists or not depending on the property of the network under consideration, however a specific characterization of community structures may guide for a good conception of detection mechanism or an appropriate choice of community detection technique. 

Specifically, we focus on the evaluation of structural communities, which means communities are distinguished based on the interaction between their nodes through edges but not on contextual information neither network meta-data. One could criticize this approach since it is also possible that real communities in networks are not structurally good but yet well cohesive according to a more natural sense of community. However, most of the time, contextual information or metadata are missing and the only way to discover community structure is using network topology. Moreover, a generic analysis using only interaction information enables a comparative approach to contrast communities throughout different network categories, which are not allowed by sophisticated approaches using contextual information. % and a matching to associate them with similar graph models.

The paper is organized in the following way: firstly, some close related work are presented in Section~\ref{sec:related} to give an overview about similar existing researches; then quality metrics that are used to characterize structural communities will be introduced in Section~\ref{sec:quality}. Then, Section~\ref{sec:methods} describes community detection methods employed to identify and validate communities; subsequently, the network dataset and their uncovered hidden communities are analyzed in Section~\ref{sec:empericalanalysis} using quality metrics presented in Section~\ref{sec:quality}. After previous introductions and analysis, we present a conceptualization of community structure using a bivariate representation approach which is introduced in Section~\ref{sec:cate}, then based on this approach we identify in Section~\ref{sec:category_profile} different structural profiles of communities in various network categories such as: communication, technological, information, biological, social, ecological and synthetic networks. Finally, Section~\ref{sec:discussion} draws some conclusions and envisages some potential perspectives for future work.

\section{Related work}
\label{sec:related}

Lancichinetti \textit{et al.} characterize community structures of complex networks in different domains by observing the evolution of various qualities such as community scaled density, average shortest path, max internal degree, etc. in large scale networks according to discovered community size ~\cite{lancichinetti:2010a}. The evolution of these qualities in function of number of nodes in each cluster helps the authors to deduce and characterize different structures found in many class of networks such as: Internet, communication, information, biological and social networks. 

Guimera \textit{et al.} demonstrate that modular networks in real world can be classified into distinct functional classes depending on the composition of connection profiles between their nodes~\cite{guimera:2007a}. Specifically, by using two metrics including within-module degree $z$ and participation ratio $P$~\cite{guimera:2004a}, a node in a community is characterized by seven different roles of hubs and non-hub nodes. Once the role of every node in a network partition is determined, the connectivity profiles of interactions in the network can be analyzed. Specifically, the authors determine two main classes of networks based on the presence of role-to-role connectivity profiles. The first class called \textit{string-periphery} includes metabolic and air transportation networks which are rich in ultra peripheral interactions and hub interactions. The second class called \textit{multi-star} includes protein interactome and Internet networks which are, on the other hand, rich in ultra peripheral-provincial hub interactions. 

Leskovec \textit{et al.} investigate the variation of community structure in large scale networks using \textit{conductance} metric~\cite{leskovec:2008a}. In fact, the authors measure the variation of the lowest community conductance in function of community size. This variation depicts a so-called \textit{network community profile} which helps to characterize community quality over a wide range of size scales. The authors also point out that communities attain the best quality (in terms of conductance) at a characteristic size of around $100$ nodes and provide evidences of a high presence of core-periphery community structure in real networks through numerous empirical experiences. 

Coscia \textit{et al.} generalize the problem of community detection discovery by reconsidering the question of what can be considered to be a community~\cite{coscia:2012a}. The authors then resume popular methods in the literature according different quality aspects such as density-based, vertex similarity-based, action-based or influence propagation-based. A definition-based classification of community discovery methods according to a large number of community features is then introduced. This classification approach shifts the attention from how communities are detected to what kind of communities to detect and provides another point of view regarding to community detection.

The most common and fundamental point between this paper and the previously mentioned work is the exploratory objective to characterize communities in complex networks by observing qualities using statistical metrics. Concretely, we contribute a methodology to describe community topologies in a systematic and generic way that can be extended to any category of networks. This means one can mechanically apply the same analysis procedure to explore community structures of any network of interest. 

\section{Community quality characterizing metrics}
\label{sec:quality}
The purpose of this work is characterizing structural properties of communities that could be found by using various community detection techniques. Since the idea of community structure varies from one context to another, it is not expected that a finite set goodness features could fit every intuition of what a good community is and the choice of any set of metrics would be adversarial unless a specific context is clearly defined under a constrained circumstance. Meanwhile, many goodness metrics define community qualities based on the conditions where they are found. Consequently, in order to remain the analysis as generic as possible, these metrics are not considered in this paper to characterize communities. Therefore, we restrict our list of quality metrics of interest in the analysis by applying the following criteria from the highest to the lowest priority:

\begin{itemize}
\item Since we are characterizing communities in different types of networks, we are only interested in metrics who delineate communities themselves, not in a relative relation with the global structure of networks where they are found (such as Cut ratio~\cite{yang:2015a}, Modularity~\cite{newman:2004a} or Description Length~\cite{rosvall:2008a}) even though their efficacy can not be ignored.

\item Potential metrics for the analysis must be relatively uncorrelated from one to another throughout a wide range of networks in order to illustrate different aspects of structural characteristics.
%\item A metric must reflect the internal structure of a community

\item A metric whose concepts can be represented intuitively and visually in order to describe most distinguishable characteristics is preferable than a metric that reflects statistical ideas which are difficult to be presented by simple topologies. 

\end{itemize}

In the later part, we show that the choice of structural metrics using these criteria helps to distinguish community structures that are found in many categories of networks using different techniques of community detection. But firstly, we introduce some general notations that helps to define metrics commonly used to measure network or community structures. 

\subsection{General notations}
We formulate different quality metrics using the following notations: given an undirected and unweighted graph $G = (V,E)$ which is composed of a set of $n = |V|$ nodes and $m=|E|$ edges where $E=(u,v):u,v \in V$. Each node in a graph is characterized by a degree $d(u)$ which is the number of connections that it has with the other nodes in the graph. Given a cluster $S$ of $n_S$ nodes, which is a subgraph of $G$, a function $g(S)$ quantifies a structural goodness feature of $S$ according to a particular expectation of community quality. We denote $m_S$ as the number of edges inside $S$, $m_S = |{(u,v) \in E : u \in S, v \in S}|$; $c_S$ be the number of edges that connect $S$ to other nodes outside of $S$,  $c_S = |{(u,v) \in E : u \in S, v \not\in S}|$. The number of connections of a node $u$ in a subgraph $S$ with other nodes in $S$ is called internal degree of node $u$ and is denoted as $d_{int}(u)$. As such, the external degree of node $u$ can be calculated as $d_{ext}(u) = d(u) - d_{int}(u)$. The relation between internal and external degrees of nodes in $S$ with its number of edges can be resumed as: $\sum_{u\in S} d_{int}(u) = 2m_S$ and $\sum_{u\in S} d_{ext}(u) = c_S$. 

Based on the previous preliminary, we present in the following section a list of commonly used goodness metrics which can be classified in four principle families:

\subsection{Metrics based on internal edge density}
\begin{itemize}
\item Density
\begin{equation}
\textsf{density}(S) = \frac{m_S}{n_S(n_S-1)/2}
\end{equation}
This metric captures the idea that nodes in a community must be densely connected wherever possible. It quantifies the fraction of edges inside $S$ over the total possible edges that could be established.

\item Scaled density
\begin{equation}
\textsf{sc\_den}(S) = \frac{2m_S}{n_S-1}
\end{equation}
Scaled density is a kind of normalized density which is defined as $n_S$ times the density of the community~\cite{lancichinetti:2010a}, \cite{labatut:2017a}. This normalization is usually applied to palliate an issue due to the fact that the number of edges in a sparse network increases linearly with its size, however the number of possible edges increase quadratically. As a consequence, traditional edge density could not well distinguish large communities and this modification by a multiplication with the number of node is believed to reflect better edge density concept in real world networks. 

\subsection{Metrics based on centralized or hub structure}
\item Hub dominance
\begin{equation}
\textsf{hub\_dom}(S) = \frac{max_{u \in S} d_{int}(u)}{n_S - 1}
\end{equation}
Internal edges of a community can be distributed in various ways around its nodes, either concentrating around a few numbers of high centralized nodes or uniformly divided into every node. The hub dominance metric is designed to identify the level of central organization around well connected nodes. The higher this metric of a community, the more likely it has a hub-like structure~\cite{lancichinetti:2010a}, \cite{labatut:2017a}. 

\subsection{Metrics based on triadic structure}

\item Clustering coefficient (CCF)
\begin{equation}
\textsf{CCF}(S) = \frac{3 \Delta_S}{T_S}
\end{equation}
Where $\Delta_S$ denotes the number of triangles in community $S$ and $T_S$ indicates the number of triples of vertices in $S$, which means number of connected subgraphs consisting of 3 vertices. This metric reflects the probability that the adjacent vertices of a vertex are connected. This is a well-known metric which is usually used to evaluate modular structure in networks. It is based on the concept that pairs of nodes with common neighbors are more likely to be connected~\cite{barrat:2004a}.

\item Triangle participation (TPR)
\begin{equation}
\textsf{TPR}(S) = \frac{\sum_{u \in S} \delta_{uS}}{n_S}
\end{equation}
Where $\sum_{u \in S} \delta_{uS} = 1$ if node $u$ belong to at least one triangle in community $S$ and $\sum_{u \in S} \delta_{uS} = 0$ if node $u$ does not belong to any triangle in community $S$~\cite{yang:2015a}. There is a slightly difference between the clustering coefficient and the triangle participation, while the former considers a good community based on the number of possible connections which could be constructed in the community, the latter only cares about whether there are many individuals of the community participate or not in tight connections (cliques). 
\end{itemize}

\subsection{Metrics based on external connectivity}
\begin{itemize}

\item Expansion: 
\begin{equation}
\textsf{expansion}(S) = \frac{c_s}{n_s}
\end{equation}
The metric measures the number of edges per node that point out side a cluster~\cite{yang:2015a}. It represents the relative out degree of a cluster over its size. The higher the expansion of a community, the stronger the its connection with the rest of the network. Generally, in a common sense, community detection methods try to minimize inter-community connectivity and hence reducing community expansion.

\item Conductance
\begin{equation}
\textsf{conductance}(S) = \frac{c_s}{2m_S+c_S}
\end{equation}
The conductance represents the fraction of degrees of a community that points outside over the total of its degrees. The conductance reveals how much the direct neighbors of a node in the community belong to neighborhood communities. In other words, the higher the conductance, the more likely that nodes connect to the community belong to another community. Leskovec \textit{et al.} show that finding a configuration in networks that minimizes the conductance of communities helps to identified good network community profile~\cite{leskovec:2008a}. 

\item Average Out Degree Fraction
\begin{equation}
\textsf{meanODF}(S) = \frac{1}{n_S} \sum_{u\in S} \frac{|(u,v) \in E: v\notin S|}{d(u)}
\end{equation}
with $ODF_S(u) = \frac{|(u,v) \in E: v\notin S|}{d(u)}$ is called the out degree fraction of node $u$ in subgraph $S$.  
The \textsf{meanODF} value indicates the average of out degree fraction of nodes in a community, a low \textsf{meanODF} implies that nodes in the community connect primarily with other nodes inside the community while a high \textsf{meanODF} means that nodes connect preferably to nodes in other communities rather than to the ones in its own~\cite{yang:2015a}. 

\item Maximum Out Degree Fraction
\begin{equation}
\textsf{maxODF}(S) = max_{u \in S} \frac{|(u,v) \in E: v\notin S|}{d(u)}
\end{equation}
The \textsf{MaxODF} reflects the maximum of fraction of edges of a node in community $S$ that connect outside $S$. This metric helps to quantify the interaction of the most active node of community $S$ with the rest of the network. 
\end{itemize}

Besides, there are many other metrics in this family such as \textsf{FlakeODF}, \textsf{Cut ratio}, \textsf{Normalized cut}, etc. However they expose high correlations with specified metrics in our analysis and are not listed here~\cite{yang:2015a}.

\section{Community detection methods}
\label{sec:methods}
In this section, we present community detection methods that have been used in order to study community structure in our network dataset. The choice of analyzing methods should not be considered neither an exhaustive nor a well representative list. Since the objective of this paper is to focus on different modular structures that could potentially be identified on real-world networks, we only chose a few numbers of methods whose performances have been proven in the literature to be reliable and can be well accessed by a large numbers of analysts. The only criterion we take into account is that these methods use different approaches to discover communities. While the edge betweenness method~\cite{girvan:2002a} is based on edge centrality detection in order to break networks into several communities; the Louvain method~\cite{blondel:2008a} optimizes local modularity by iteratively folding nodes into meta-nodes; the label propagation~\cite{raghavan:2007a} determines the community of a node by considering the memberships of its neighbors; and the Infomap method~\cite{rosvall:2008a} relies on finding a configuration that maximizes the compression of a random walks represented by an encoded binary sequence. Of course one could argument that by using only a few numbers of methods, it is likely that some kind of structures are not well covered in the analysis. Although it is a very pertinent requirement, within this study, the authors find that the utilization of some representative methods could already help to reveal substantially many interesting community structures. A summary of community detection methods that have been used to discover networks is illustrated in Table~\ref{tab:methods}.

\begin{table}[h]
\centering
\begin{tabularx}{0.5\textwidth}{l X r}
\hline
\textbf{Method} & \textbf{Approach} & \textbf{Complexity}\\
\hline
Edge betweenness~\cite{girvan:2002a}& Betweenness centrality & $\mathcal{O}(nm^2)$ \\
Fast greedy~\cite{aaron:2004a} & Modularity & $\mathcal{O}(nm\log n)$ \\
Louvain~\cite{blondel:2008a} & Local modularity & $\mathcal{O}(\log n)$ \\
Spectral ~\cite{newman:2004a} \hspace{5pt} & Modularity & $\mathcal{O}(n(m + n))$ \\
Walktrap~\cite{pons:2005a} & Dynamic distance & $\mathcal{O}(mn^2)$ \\
Label propagation~\cite{raghavan:2007a} & Topological closeness & $\mathcal{O}(m + n)$ \\
Infomap~\cite{rosvall:2008a} & Information compression& $\mathcal{O}(n(m + n))$ \\
Spin glass~\cite{reichardt:2006a} & Energy model & $\mathcal{O}(n(m + n))$ \\
\hline 
\end{tabularx}
\caption{A summary of community detection algorithms used to study community structure in the analysis. They are used to identify communities in the network dataset.}
\label{tab:methods}
\end{table}

\subsection{Edge betweenness community detection}
Girvan and Newman proposed a method~\cite{girvan:2002a} to identify boundaries between communities in network by measuring a factor called \textit{edge betweenness centrality} which reveal the contribution of each edge in the network for constructing shortest path way between two arbitrary nodes. An edge with a high \textit{betweenness centrality} indicator represents an important connection that joints two \textit{compact} groups in a network. It means that it is very likely that the shortest pathway between two nodes of these groups must go through the edge. Thus, removing such a high \textit{edge betweenness} degree connection will probably separate two loosely connected clusters. The authors construct a community detection method based on the idea that if one gradually removes high \textit{betweenness centrality} edges in a network, after some iterations, the network will be disconnected and nodes located in different components can be considered as prospective communities. 

\subsection{Fast greedy method}
This method discover communities in networks using a greedy method to optimize the modularity objective function throughout many iterations. The modularity $Q$ is defined as the difference between the number of edges within communities and the number of expected number of such edges. It can be written: $Q = \frac{1}{2m} \sum_{ij} [A_{ij} - P_{ij}] \delta(S_i,S_j)$, where $P_{ij}$ represents the probability of having an edge between $i$ and $j$, $\delta(S_i,S_j) = 1$ if $S_i = S_j$ and $\delta(S_i,S_j) = 0$ otherwise. This is a hierarchical agglomeration algorithm which is well-known for its competitiveness in time complexity $\mathcal{O}(nm\log n)$ and can be reduced to $\mathcal{O}(n\log^2n)$ in sparse networks where $n$ corresponds to the number of vertices of the network~\cite{aaron:2004a}. The idea of the authors to reduce the complexity of the agglomeration process is only that they only consider amalgamations between nodes that share at least one common edge. In that way, they can make use of the data structure of the original graph in order to keep track of the modularity changes $\Delta Q_{ij}$ when merging two nodes $i$ and $j$.       

\subsection{Multi-level community detection - Louvain method}
This heuristic method employs modularity optimization in order to discover hierarchically community structure for networks and is claimed to outperform all other known methods in terms of computation time~\cite{blondel:2008a} with a good compromise quality measured by modularity on large networks. The algorithm is executed through two concatenated phases that repeat iteratively. At the first step, each node of the network belongs to its own community and is considered to be merged with its neighbors to establish a new community in a way to gain a maximum increase of modularity. The second step consists of constructing an aggregated network where communities in the original network become new nodes; links between two new nodes are given by the sum of the weight of the links between nodes in the corresponding two communities; links between nodes inside a same community in the first step become self-loop links in the new network. Once the second phase is finished, the first phase of the algorithm is then reapplied to the resulting network. These two phases are iterated until no additional modularity is obtained.         

\subsection{Walktrap method}
This method estimates the distances between vertices in a network using a structural similarity measure based on random walks. Such that the distance between two vertices must be large if they belong to different communities and must be small if they belong to the same community. The approach relies on the intuition that a random walker may have a tendency to be trapped into densely connected parts of networks where nodes have a similar stochastic state. In fact, two vertices are considered to be similar and belong to the same community should ``see'' the other vertices in the same way. Specifically, the authors proposed to define a similarity distance $r_{ij}$ between two vertices $i$ and $j$ as a function of the difference between the probabilities $P_{i \cdot}^t$ and $P_{j\cdot}^t$ to go from $i$ and $j$ to other vertices in a short number of $t$ steps: $r_{ij}^2 = \sum_{k=1}^n \frac{(P_{ik}^t-P_{jk}^t)^2}{d(k)}$~\cite{pons:2005a}. Then, a traditional hierarchical clustering algorithm is used to find communities from this dynamic distance.

\subsection{Label propagation method}
This method is based on the idea that nodes in a network have a tendency to participate into the community where the majority of their neighbors are found~\cite{raghavan:2007a}. The algorithm initializes every node with a unique label and repeats modifying these labels at every iterative step. In each step, each node adopts a new label that most of its neighbors currently have and this process is expected to help identifying densely connected groups of nodes that have unique labels and considered as communities. Ideally, the iterative process should converge when no node in the network changes its label, however it is normally possible that nodes in a network have an equal maximum number of neighbors in two or more communities. In these cases, the algorithm breaks ties randomly among the possible candidates.

\subsection{Infomap method}
With a primal purpose to understand the flow of information on networks, this method is designed in order to decompose a network into communities by optimally compressing description of information flows on the network~\cite{rosvall:2008a}. There is a conceptual distinction of community notion in this method with the traditional ones. In fact, instead considering density-related elements, the authors consider a community as a group of nodes among which information flows quickly and easily, and so that they can be aggregated and represented by a single well-connected module. In order to do that, it can be imagined that the algorithm employs a random walker to describe information flows on a network of interest and then exploits the regularity of the random walker's path that have been traced and encoded. The modules are then determined as the configuration that minimizes the amount of necessary codeword length in order to compress the regularity in the path of the random walker.

\subsection{Modularity spectrum-based method}
This method searches for a partition that maximizes the modularity fitness function using a spectral partitioning calculation on a \textit{modularity matrix} \textbf{$B$}~\cite{newman:2006a}. This matrix is defined as $B_{ij} = A_{ij} - P_{ij}$ where $A_{ij}$ denotes the adjacency matrix of the graph under consideration and $P_{ij}$ represents the probability $p_{ij}$ for and edge to fall between the pair of vertices $i,j$. Here, $P_{ij}$ reflects the expectation of the existence of an edge between two arbitrary vertices in an associated graph where node degree sequence keeps unchanged and modular structure is considered not to be presented. Inspired from the spectral partitioning problem to minimize the cut size of a clustering using $Laplacian$ matrix, the author find out that the leading eigenvectors that correspond to the positive eigenvalues of the \textit{modularity matrix} helps to find a good partition that maximize the modularity. 

\subsection{Spin glass method}
In this method, the problem of community detection is interpreted as finding the ground state of  an infinite range spin glass, which is the configuration that minimizes the energy of the system. Determining this state configuration suggests useful information to locate communities being groups of nodes that have the same states. The basic principle of the model is that edges should only connect vertices of the same class, which have the same spin state. Here, the formulation of system energy at the same time rewards internal edges between nodes of the same group and penalizes missing edges between nodes in the same group, penalizes edges between nodes of different groups and rewards non-links between different groups~\cite{reichardt:2006a}.

\section{An empirical analysis}
\label{sec:empericalanalysis}
In this section, we describe some statistical properties of networks that will be included in the following analysis. It is expected that networks in each category are spread in a wide range of structural measures. However, available biological networks that have been published and analyzed widely are relatively small in comparison to the other networks of the other families. Besides, due to the complexity of the analysis process, we limit the domains of interest at 5 categories which are commonly researched and where numerous networks are available. The number of networks considered is 108 which is relatively large in comparison to many studies in the art. Many notable related work where some of these networks are also employed to study community structure could be mentioned for a quick reference: Orman \textit{et al.} use 6 networks to evaluate the structure of communities discovered by several detection techniques~\cite{orman:2012a}; Lancichinetti \textit{et al.} use 15 networks to characterize structural communities~\cite{lancichinetti:2010a}; Darko \textit{et al.} use 16 networks to reveal differences between structural communities and ground truth; Leskovec \textit{et al.} use over 100 networks to analyze network community profile~\cite{leskovec:2008a} and 230 networks to evaluate the goodness of ground-truth communities in social networks, within this number, 225 samples of the Ning platform's networks~\footnote{https://www.ning.com/} are aggregated~\cite{yang:2015a}. Table~\ref{tab:dataset} resumes the composition of networks that have been analyzed in this paper. 

\subsection{A description of empirical network dataset}
\begin{table}[h]
\centering
\begin{tabularx}{0.5\textwidth}{l X X X r}
\hline
\textbf{Category} & \textbf{Size} & \textbf{Nodes} & \textbf{Edges} & \textbf{Notable networks} \\
\hline
Biological & 7 & 1860 & 10763 & Protein, yeast\\
Communication \hspace{5pt} & 9 & 39595 & 195032 & Email, forums\\
Information & 25 & 38358 & 159812 & Citation, Amazon\\
Social & 37 & 6888 & 49666 & Facebook, Youtube\\
Technological & 19 & 18431 & 48494 & Internet, P2P\\
Miscellaneous & 11 & 4298 & 49033 & Ecology, synthetic\\
\hline
Total$^*$ & 108 & 1.99M & 9.08M & \\
\hline
\end{tabularx}
\caption[A summary of network dataset used in this analysis where \textbf{Size} is the number of networks analyzed in each category, \textbf{Nodes} and \textbf{Edges} indicates the average number of nodes and edges of networks in each category respectively. $^*$The last row shows the total number of networks, nodes and edges in the whole dataset.]{A summary of network dataset used in this analysis where \textbf{Size} is the number of networks analyzed in each category, \textbf{Nodes} and \textbf{Edges} indicates the average number of nodes and edges of networks in each category respectively. $^*$The last row shows the total number of networks, nodes and edges in the whole dataset. This dataset is collected from several sources including: \url{http://networkrepository.com} \protect{\cite{rossi:2015a}}, \url{http://konect.uni-koblenz.de}~\protect{\cite{jerome:2013a}}, \url{http://snap.stanford.edu}~\protect{\cite{snapnets}}}.
\label{tab:dataset}
\end{table}

\begin{figure*}
\centering\includegraphics[width=0.85\linewidth]{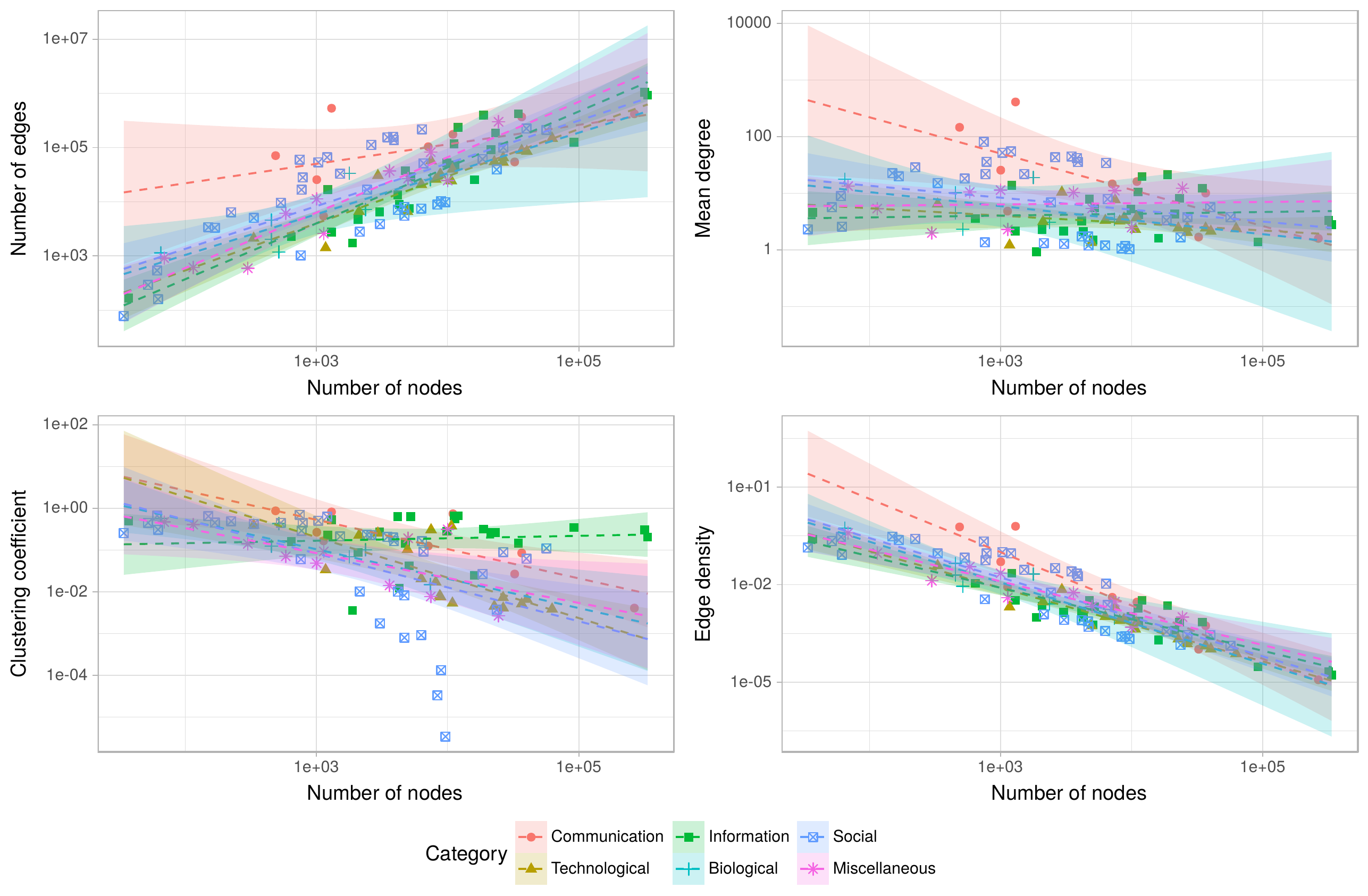}
\caption{From left to right, up to bottom, we illustrate structural measures of the 108 networks: (a) Number of edges as a function of the number of nodes, (b) Mean degree $\langle k \rangle$ as a function of the number of nodes, (c) Clustering coefficient in function of the number of nodes, (d) Edge density in function of the number of nodes. The dark backgrounds represent the $95\%$ confidence intervals of the regression model outputs used to estimate the linear relations for each network category.}
\label{fig:network_dataset}
\end{figure*}

Some notable structural measures of networks in the dataset are illustrated in Figure~\ref{fig:network_dataset}. It is noticeable that apart from biological networks which are relatively small, the other classes cover quite a wide range of number of nodes, edges, mean degree, clustering coefficient and edge density. Since real world networks are relatively sparse, the number of edges increase linearly in function of the number of nodes and consequently, the edge densities decrease linearly by number of nodes (since the number of possible connection increase quadratically by number of nodes). This sparsity property can be easily noticed from Figure~\ref{fig:network_dataset}(a,d). From Figure~\ref{fig:network_dataset}(b), it can be seen that the mean degrees of the networks in the dataset vary principally between $1$ and $100$ edges per node except for 2 communication networks. In a global point of view, networks in the dataset have a quite strong modular quality since most of them have relatively high clustering coefficient as shown in Figure~\ref{fig:network_dataset}(c). This quality will be investigated more in community-level in a following section.

\subsection{Evaluating community structures using quality metrics}
In other to characterize structural communities of different types of networks, we apply various community detection methods on the dataset grouped by category of networks. Once communities are produced, the quality metrics  presented in Section~\ref{sec:quality} are used to evaluate the quality of detected communities. Since many metrics reflect close structural properties, we analyze the correlations between the corresponding qualities on the detected community sets. This analysis allows to select only the most representative structural metrics to delineate community structures.    

\begin{figure*}
\centering\includegraphics[width=1\linewidth]{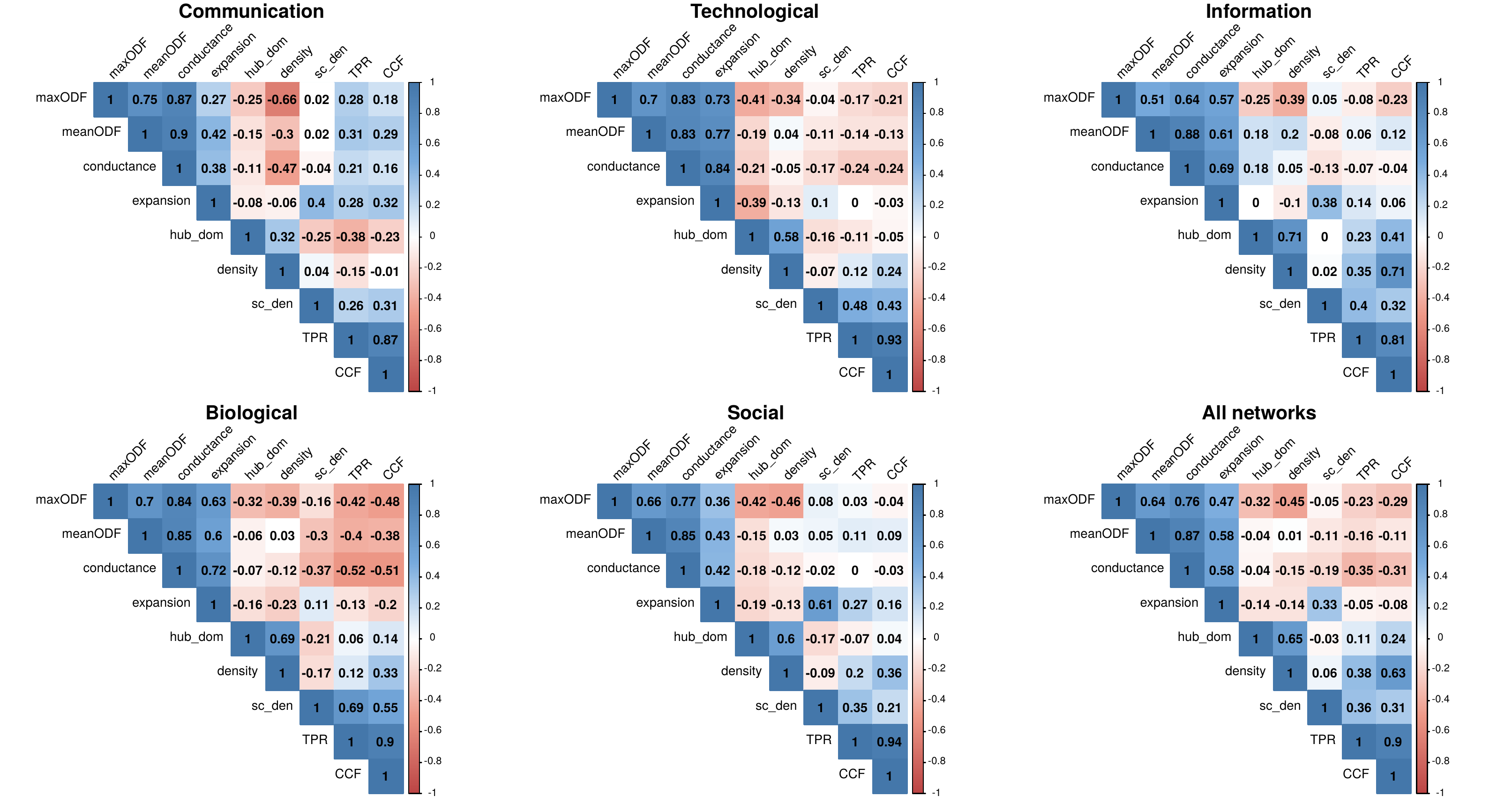}
\caption{The Pearson correlations of community metrics measured on the communities detected by the set of community detection methods on the network dataset. These correlation are calculated based on scores of metrics measured on communities that contain at least 3 nodes. Metric correlations are analyzed by group of networks in different domains. Quality metrics are presented in the 6 sub-figures in the same order for a comparative observation. Correlation scores with low estimated significant levels ($P$-value $> 0.01$) are reproduced in a blank background.}
\label{fig:corr_kpi}
\end{figure*}

Figure~\ref{fig:corr_kpi} illustrates the correlation matrices of different structural qualities measured on various community sets identified by community detection algorithms presented in Section~\ref{sec:methods} over 5 classes of networks and the whole dataset. Note that only communities whose sizes are at least 3 nodes are taken into consideration in the figure since many metrics are meaningless for too small communities (which contain one or two nodes). It is important to note that although some statistical metrics are only significant when measuring on large communities, the corresponding correlation matrices for large scale communities resemble globally with those of Figure~\ref{fig:corr_kpi}. Specifically, a calculation using only large communities of more than 10 nodes gives quite similar and consistent correlation scores. The employment of representative quality metrics is globally justifiable on the whole range of community size scales. And so that the same metrics can be relatively significant to represent communities on the whole range of community size.  

As we can see in Figure~\ref{fig:corr_kpi}, there are two groups where metrics are consistently correlated from one to another. The first group includes \textsf{maxODF, meanODF} and \textsf{conductance} who represent community external connection with very high correlation coefficients (except for \textsf{maxODF} and \textsf{meanODF} in information networks with a relatively weak relation of $0.51$). Besides, the \textsf{expansion} metric also belong to this group in technological, information and biological networks with high correlation scores and more loosely in the other types of networks. The second group consists in \textsf{TPR} and \textsf{CCF} who expose triadic tight-knit structures and are observed with very high correlation scores in every case of network category. The lowest correlation score between \textsf{TPR} and \textsf{CCF} is reported at $0.81$ in information networks and approximately around $0.90$ in all the other cases. Without loosing the generality, in our analysis, these 2 groups of metrics could be reduced to two representative metrics representing two structural properties. 

Hub dominance (\textsf{hub\_dom}) is the only metric who is quite independent of all metrics in the two previous groups in every network category. The highest absolute correlation score between \textsf{hub\_dom} with these metrics is $0.42$ with \textsf{maxODF} in social networks, which is still a relative low correlation. This latter, however, is generally correlated with \textsf{density} except for the case of communication networks where they are quite orthogonal. In the mean while, scaled density (\textsf{sc\_den}) shows an inconsistent association throughout the studied network categories. It is close to \textsf{CCF} and \textsf{TPR} in biological networks but approaches \textsf{expansion} in social networks. 

Based on this analysis, the above community quality metrics can be grouped in 6 classes that are presented in Table~\ref{tab:dimension} according to their correlations over the studied dataset. In other words, these quality metrics are more correlated with ones in the same groups than with the others. Consequently, it is preferable to describe community structure using a cross combination of metrics in these groups. We present in the following section a characterization of internal community structure by a descriptive approach using an association between metrics in 2 different groups. Then we demonstrate by empirical evidences that our approach helps to recognize different community structures in communication, information, technological, biological, social, ecological and synthetic networks. 

In fact, the previous analysis shows that internal and external structures of communities are generally not correlated. They reflect different facets of community structures. Consequently, the characterization of community structure can be realized separately from 2 distinguishable levels of observation. In this paper, we focus on characterizing internal community structure. Readers who are interested in analyzing community external connectivity can refer to another work presented in~\cite{dao:2017a} where communities are portrayed by two variables: the level of external interaction and the distribution of these interactions over community border nodes.

\begin{table}[h]
\centering
\begin{tabularx}{0.47\textwidth}{l r}
\hline
\textbf{Metrics} & \textbf{Common concept} \\
\hline
\textsf{maxODF,meanODF,conductance} & External activeness\\
\textsf{expansion} & External connectivity \\ 
\textsf{hub\_dom} & Centralized connectivity \\
\textsf{density} & Internal edge density \\
\textsf{sc\_den} & Average internal density \\
\textsf{CCF, TPR} & Internal triadic closure\\
\hline
\end{tabularx}
\caption{Groups of quality metrics that reflect different aspects of community structural property. Two metrics belong to a same category if they show a high correlation over the sets of structural communities. The \textbf{Common concept} column precises common structural features that members of each group reflect.}
\label{tab:dimension}
\end{table}

\section{A bivariate characterization of community structure}
\label{sec:cate}
In this part, we present a categorization of community structure in a descriptive way to illustrate different modular structures. This is an extension of our previous proposition in evaluating communities using a descriptive approach~\cite{dao:2017a} for internal aspect of community structure. We propose a categorization of modular structures using a couple of representative goodness variables to reflect highlight structural characteristics of communities in real world networks. Here, we focus on internal community structure, i.e. \textsf{density, sc\_den, CCF, TPR} and \textsf{hub\_dom}, will be in the shortlist of interest.

\subsection{Which metrics fit?} 
It is well-known that \textsf{density} have a weakness in describing communities of different sizes since in real networks, the number of edges normally increases linearly with its size (real networks are often sparse) but the number of possible connection increases quadratically. As a consequence, the quality of large communities is usually under evaluated in comparison to small communities. Scaled density (\textsf{sc\_den}) palliates this issue by multiplying the density with the community size, so mathematically its concept is very close with the average degree of a community which is measured by $<k_S> = \frac{2m_S}{n_S}$. This metric reflects a very important feature of communities and is often used to evaluate community quality in a common sense. However, given a specific value of scaled density, one have several ways to redistribute edges inside a community in a manner that its internal topology changes crucially. In other words, scaled density does not characterize community internal configuration of degrees. This is the reason why we do not use scaled density or traditional density to represent community topology.

The clustering coefficient and the triangle participation ratio (\textsf{CCF} and \textsf{TPR} respectively) are relatively close in their definition and it has been proved to be highly correlated through the previous empirical analysis. They reflect an important topological feature by implying the concept that two arbitrary neighbors of a node in a community should be also connected. This idea is somehow relatively close with the density qualities since a network with high \textsf{CCF, TPR} scores is normally dense; however the opposite way is not always correct, which means a dense network does not necessarily have many triangular connections. Here, we select one metric among \textsf{CCF} and \textsf{TPR} to describe a common structural property called \textit{transitivity}. Depending on the topology of networks or communities under consideration, one metric will work better than the other. On a same network, \textsf{CCF} score is generally lower than \textsf{TPR} score and hence \textsf{CCF} has a better resolution for networks where triangles are dense. On the the other side, \textsf{TPR} magnifies better topological differences in networks where only a few triangles exist. A further investigation on the dataset shows that there is approximately $90\%$ of networks whose clustering coefficients are larger than $0.01$ and this number is around $60\%$ for a coefficient of $0.1$ (see Figure~\ref{fig:network_dataset}(d)). This evidence leads to a preference of \textsf{CCF} over \textsf{TPR} to describe the clique dominance characteristic since the networks of interest are quite dense.

Another topological dimension that we employ to describe communities is \textit{hub dominance} which is represented by \textsf{hub\_dom} metric. Similarly to \textsf{CCF} and \textsf{TPR}, this metric reflect a structural feature of edge organization in a network or community. Specifically, it characterizes whether edges are distributed around one or a few members of their community and make them becoming hubs of connection. We illustrate in the next section that the combination two dimensions quantified by a couple of values (\textsf{CCF, sc\_den}) reveals distinctive topological structures that could help to get insights on how communities in different networks look like. 

\subsection{Locating community structures in a bi-dimensional space}
After choosing two principle characteristics corresponding to two dimensions of community quality space, we describe internal community structures in different locations of this space. In order that the distinction of representative topologies in different coordinates stays clear, we profile them in a coarse-grained description level. Specifically, we considerate 4 fundamental coordinated zones corresponding to 4 underlying topologies which are emphasized in Table~\ref{tab:types}. These classes of topologies could be explained as following: 

\begin{table}[h]
\centering
\begin{tabularx}{0.49\textwidth}{l X X r}
\hline
\textbf{Type} & \textbf{Transitivity} & \textbf{Hub dominance} & \textbf{Topology}\\
\hline
1 & Low & Low & String-based \\
2 & High & Low & Grid-based \\
3 & Low & High & Star-based \\
4 & High & High & Clique-based \\
\hline
\end{tabularx}
\caption{Four distinctive topologies characterized by \textit{Transitivity} (\textsf{CCF}) and \textit{Hub dominance} (\textsf{hub\_dom}). There is no clear boundary between high and low values in the two dimensions, it is to be specified in accordance with the context. The distinction is more clear for medium and large size communities.}
\label{tab:types}
\end{table}

\begin{figure}
\centering\includegraphics[width=1\linewidth]{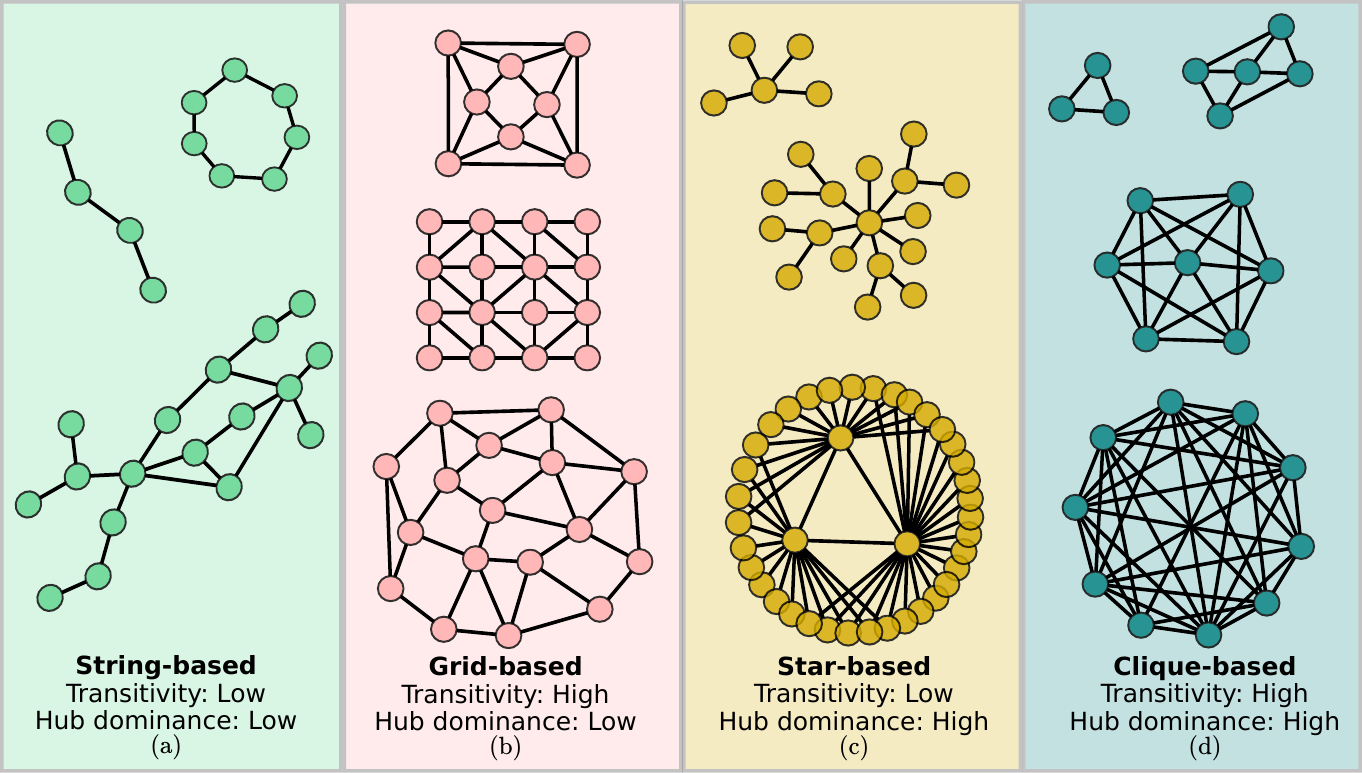}
\caption{Topology families, from left hand side to right hand side (a) String-based, (b) Grid-based, (c) Star-based, (d) Clique-based. Depending on the context, one community can belong to different topological families according to specific criteria of analyst reflected by their determination of frontiers between these families.}
\label{fig:topology}
\end{figure}

\begin{itemize}
\item \textit{String-based topology} of a community is determined by low values of transitivity and hub dominance metrics. The low scores in these two representative dimensions regulate that there is relatively nearly no presence of clique structure nor hub node. For large communities, there could be one or a few hubs and cliques established, but not enough to dominate the global structure. These communities can be considered as a consequence of a ramification between several sub-strings which generate a few loops and hubs in their intersections. String-based topologies could have a form that looks like chains, braids, rings, etc. as shown in Figure~\ref{fig:topology}(a) depending on the context.

\item \textit{Grid-based topology} can be recognized by high values of transitivity and low values of hub dominance metric. The absence of hub nodes in the community organization is probably the most common feature with the string-based topology. Hence there is a homogeneity in the connection pattern between nodes of the grid-based topology. Besides, a high value of transitivity imply that the majority of nodes participate in tight-knit triangular structures which could themselves, at the same time, be attached between one to another to create larger and compacted structures. Grid-based communities generally have large sizes since small ones are usually degenerated into strings, loops or hub structures. In other words, grid-based structures are not recognizable by observing in a small scale or a local scale of communities. Popular topologies of this family consist of lattice topology, partially mesh topology as shown in Figure~\ref{fig:topology}(b).       

\item \textit{Star-based topology} which sometimes can be considered as tree-based topology is probably one of the most popular structures in networks of many fields. It can be perceived by low values of transitivity and high values of hub dominance. A low transitivity indicates that there is not or very few cliques. On the other hand, a high hub dominance value implies the occurrence of a ``key connection" which attracts many edges in its community to become a hub. Some popular topologies which could be found in this class include: flake structure with one central hub and several peripheral hubs; hierarchical tree structure. There is actually a close relation between star-based/tree-based and string-based topology such that in some contexts, a hierarchical tree could be seen as a string and vice versa depending on the point of view. The essential difference of these two topologies which can be observed from our representation space is that the more \textit{edge-attractive} the hub(s) in a community, the more it approaches the \textit{star-based} topology. Note that in graph theory, a tree is an acyclic connected graph. However, in this context, trees accompanied by a few loops are classified in star-based topology unless loops dominate excessively the global community structure. Some representative star-based topologies are shown in Figure~\ref{fig:topology}(c). 

\item \textit{Clique-based topology} is quite common in small and very small communities but very rare in medium and large communities. It is recognized by high scores of transitivity and hub dominance. A simple interpretation of this class of topology is that every node must be connected with every other node of its community in an ideal situation. In a more relaxed context, nodes are not required to connect with all other nodes, but with a majority in order to establish a tight and compact structure. The clique-based topology is quite close to the grid-based topology in many ways. The most notable difference between them is that in a clique-based community, every node must be in the neighborhood of the other nodes of the community (direct connection or by one/two intermediate connections maximum), whether it is not necessary that every node must be close to each other in grid-based topology. Some representative clique-based topologies are shown in Figure~\ref{fig:topology}(d). 

\end{itemize}

A community structure whose transitivity and hub dominance scores are medium needs more investigation to be deduced. Since neither hub, clique nor random structure could dominate the whole community, its topology depends on the distribution of hubs and cliques in the community. It can be composed of a mixture of different component structures presented previously to become a  homogeneous and more complex topology. It can also be a simple attachment between various dissimilar structures to establish a heterogeneous unit. In a point of view of dynamic community's evolution, communities in this class might be considered as being in a transition period between elementary structures. Alternatively, it could be a saturated state where communities attain a certain diversity and remain their complex structures. Further extent researches, which will not be mentioned in this paper, are deserved to cover more exhaustive aspects of this subject   

\subsection{Locating network models in the topological space}
Based on the idea that real networks and communities are constructed throughout different mechanisms, their topologies could be in some ways mimicked by using graph generating models. We attempt to locate networks created by popular graph models of the literature in the presented space in order to match them with the most resembling representative topology.

\begin{figure}
\centering\includegraphics[width=1.00\linewidth]{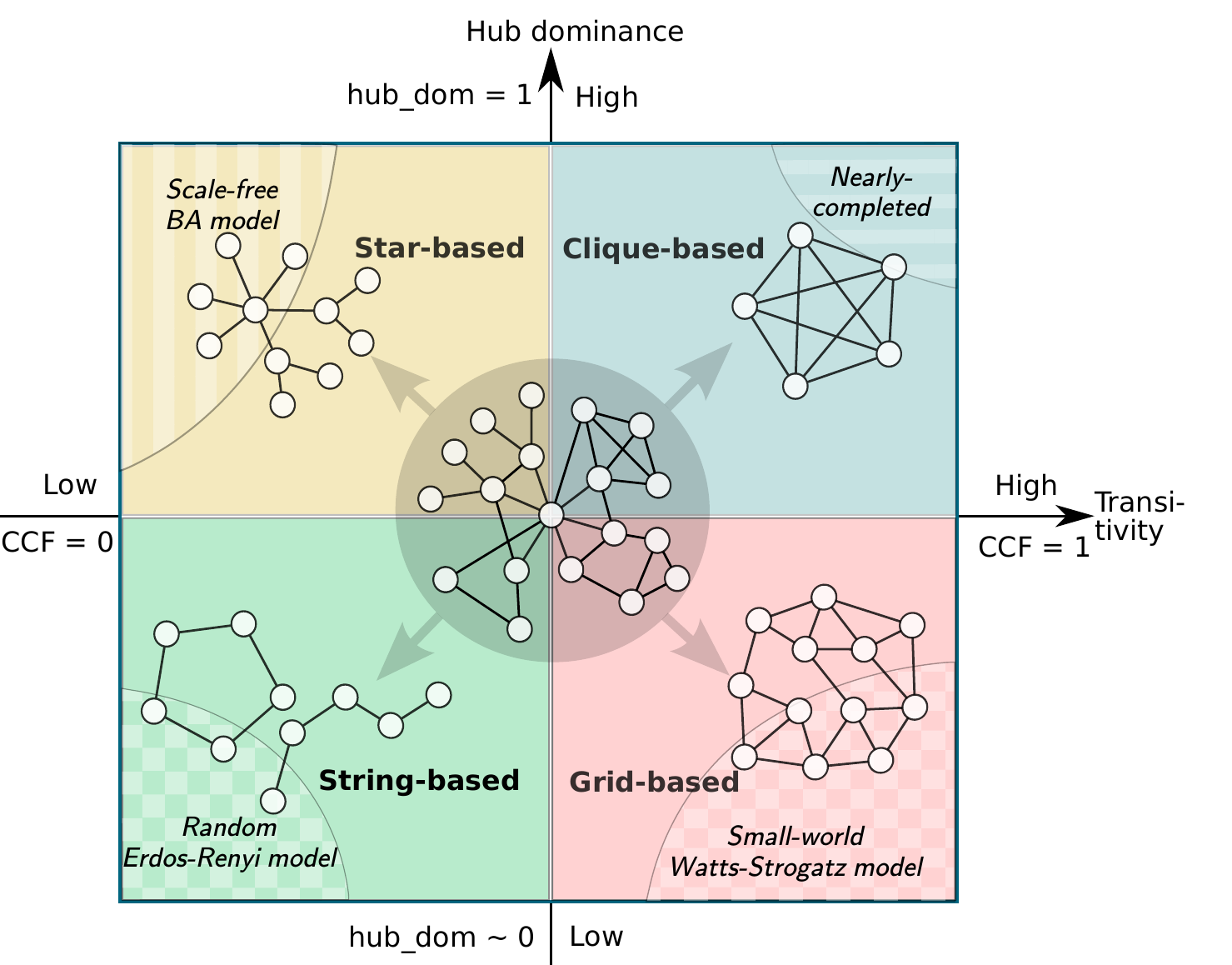}
\caption{A categorization of internal community structure according to two structural property dimensions: hub dominance and transitivity represented by \textsf{hub\_dom} and \textsf{CCF} respectively. Four representative topological communities are exemplified in 4 coordinating zones according to their corresponding \textsf{(hub\_dom,CCF)} scores. The borders between different topologies are usually not clear and can be delineated according to the context. Characteristic community size should be taken into consideration when separating characterized zones since the bigger the community size, the more likely that hubs and cliques become less significant, which means lower thresholds will be more plausible.}
\label{fig:category}
\end{figure}

\begin{itemize}
\item \textit{Erd\H{o}s-R\'{e}nyi model}\cite{erdos:1959a} is among the first models proposed to describe the generation of \textit{random graphs}. In this models, two parameters are required to generate a graph which is a fixed number of vertices $n$ and a connection probability $p$ between two arbitrary vertices (alternatively the number of edges $m$). Each pair of vertices is then connected independently of the other pairs with the probability $p$, which reflect the randomness property of the resulting graph. The expected number of edges and mean degree of the graph is calculated by $\left\langle m \right\rangle = \frac{pn(n-1)}{2}$ and $\left\langle k \right\rangle = p(n-1)$ respectively. The distribution of degree is binomial or Poisson for large graphs~\cite{newman:2001a}. If we set $n$ and $p$ parameters of the model in a way that the model creates a random graph whose average degree approaches real networks: $\left\langle k \right\rangle = p(n-1) = c > 1$, where $c$ is a constant and $c \ll n$; the graph will almost surely have a big component containing a large portion of vertices and very small components of less than $\mathcal{O}(\log(n))$ vertices. This configuration produces vertices that have all around $c > 1$ connections. In this paper, without any further mentions, we refer to random graphs as ones created by this configuration, whose average node degrees approach those of real networks. Since a random network is constructed from a homogeneous stochastic mechanism, there is normally no hubs nor cliques which means low transitivity and low hub dominance values. A typical random graph constructed with a small value of $p$ will have its largest component topology resembles the string-based topology as shown in Figure~\ref{fig:topology}(a). In an extreme regime, when the probability of connection $p$ approaches $1$, the associated random graph becomes \textit{nearly complete} as the average degree $<k>$ approaches $n-1$, which means every vertex connects with almost every other vertex as illustrated in Figure~\ref{fig:topology}(d). The location of typical random graph's topology in function of two dimensions: transitivity and hub dominance is illustrated in Figure~\ref{fig:category} in the bottom left-hand conner which associates to low scores of \textsf{CCF} and \textsf{hub\_dom}.      

\item \textit{Watts-Strogatz model} produces networks with \textit{small-world} property, which normally means that any arbitrary pair of nodes can be connected through a small number of intermediate nodes and the average geodesic distance grows proportionally to the logarithm of the number of nodes $n$ of the network: $L \propto \log (n)$. The model is built to characterize the observation that many real world networks show this property of small path length connectivity and highly clustered like regular lattices which implies a high presence of triadic closures~\cite{watts:1998a}. The generation of a small world network can somehow be considered as an interpolation between regular pattern networks and random networks. From a ring lattice with $n$ nodes and $k$ edges per node, each edge is redistributed randomly with a probability $0 < p < 1$. The authors find that a small value of $p$ reduce significantly the path length characteristic of a regular network where nodes are only connected locally. This can be explained as rewired edges create shortcuts between remote areas of the network and hence reduce considerably network characteristic distance. A typical small-world network can be described using an intermediate value of $p$, so that the distance of two arbitrary nodes are very small, the clustering coefficient stay high since the random perturbation is not strong enough to break the local structures of nodes in the lattice ring. Besides, the shape of the degree distribution in the network is quite similar to that of a random graph where every node has around $k$ neighbors and there is normally no hub dominance phenomenon. The topology of a typical small-world network is relatively homogeneous and looks like a grid-based topology from a local observation as shown in Figure~\ref{fig:topology}(b). The location of its topology in function of two dimensions: transitivity and hub dominance is illustrated in Figure~\ref{fig:category} in the bottom right-hand corner which associates to high \textsf{CCF} scores and low \textsf{hub\_dom} scores.   

\item \textit{Barab{\'a}si-Albert (BA) model}~\cite{barabasi:1999a} is originated from a discovery that the distribution of vertex degrees in many real world networks such as: genetic networks and World Wide Web networks, are quite heterogeneous. Specifically, vertex connectivity follows a \textit{power-law distribution}, which means the probability that a vertex connecting to $k$ neighbors in its network equals $p(k) = Ck^{-\alpha}$ where the constant $C$ is fixed by a normalization requirement and $\alpha$ is the power-law coefficient. This coefficient varies between $2$ and $3$ in many networks where the degree sequences are estimated to follow this model. Networks possessing this statistical feature are called~\textit{scale-free} by Barab{\'a}si \textit{et al.} to highlight the scale invariance property. This feature is explained by the authors as a consequence of two principle mechanisms: firstly, networks expand gradually by attracting new vertices to existing ones; secondly, these new vertices have a tendency to attach preferentially to vertices that are already well connected. That is why this model is often known as preferential attachment model, implying that the more connected a vertex, the more likely it receives new edges. This mechanism makes scale-free networks hub-profuse since \textit{``richer nodes get richer"}, and hence hub dominance values of scale-free networks are usually high. On the other hand, the associated clustering coefficients are usually low and are decayed quickly in function of network sizes~\cite{klemm:2002a,fronczak:2003a}, which means low transitivities. Consequently, typical scale-free networks have a close structure with that of star-based topologies as depicted in Figure~\ref{fig:topology}(c). The location of scale-free networks in function of two dimensions: transitivity and hub dominance is illustrated in Figure~\ref{fig:category} in the top left-hand corner which associates to low \textsf{CCF} and high \textsf{hub\_dom} scores.   

\end{itemize}

%\subsection{A matching between community structures and graph models}

\section{Identifying community profile of different network categories}
\label{sec:category_profile}
\begin{figure}
\centering\includegraphics[width=1.00\linewidth]{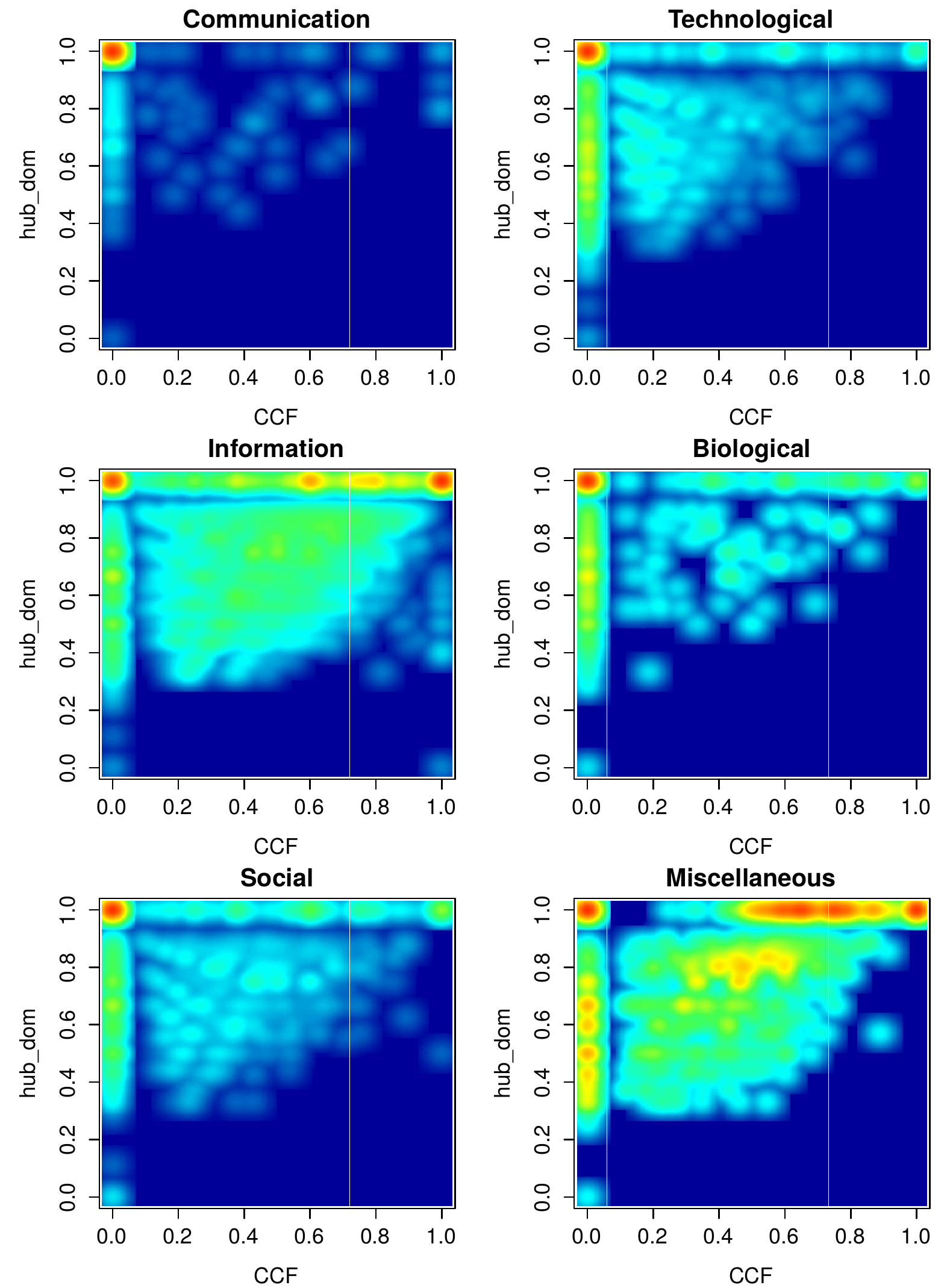}
\caption{Heat maps of distributions of small structural communities detected on different categories of networks are presented on a two dimensional space characterized by transitivity (CCF) and hub dominance (hub\_dom). Only communities of \textbf{10 nodes or less} are included. From left to right, top to bottom (a) Communication, (b) Technological, (c) Information, (d) Biological, (e) Social, (f) Miscellaneous consists in power networks, ecological networks, artificial networks, etc. }
\label{fig:ccf_hub_small}
\end{figure}

\begin{figure}
\centering\includegraphics[width=1.00\linewidth]{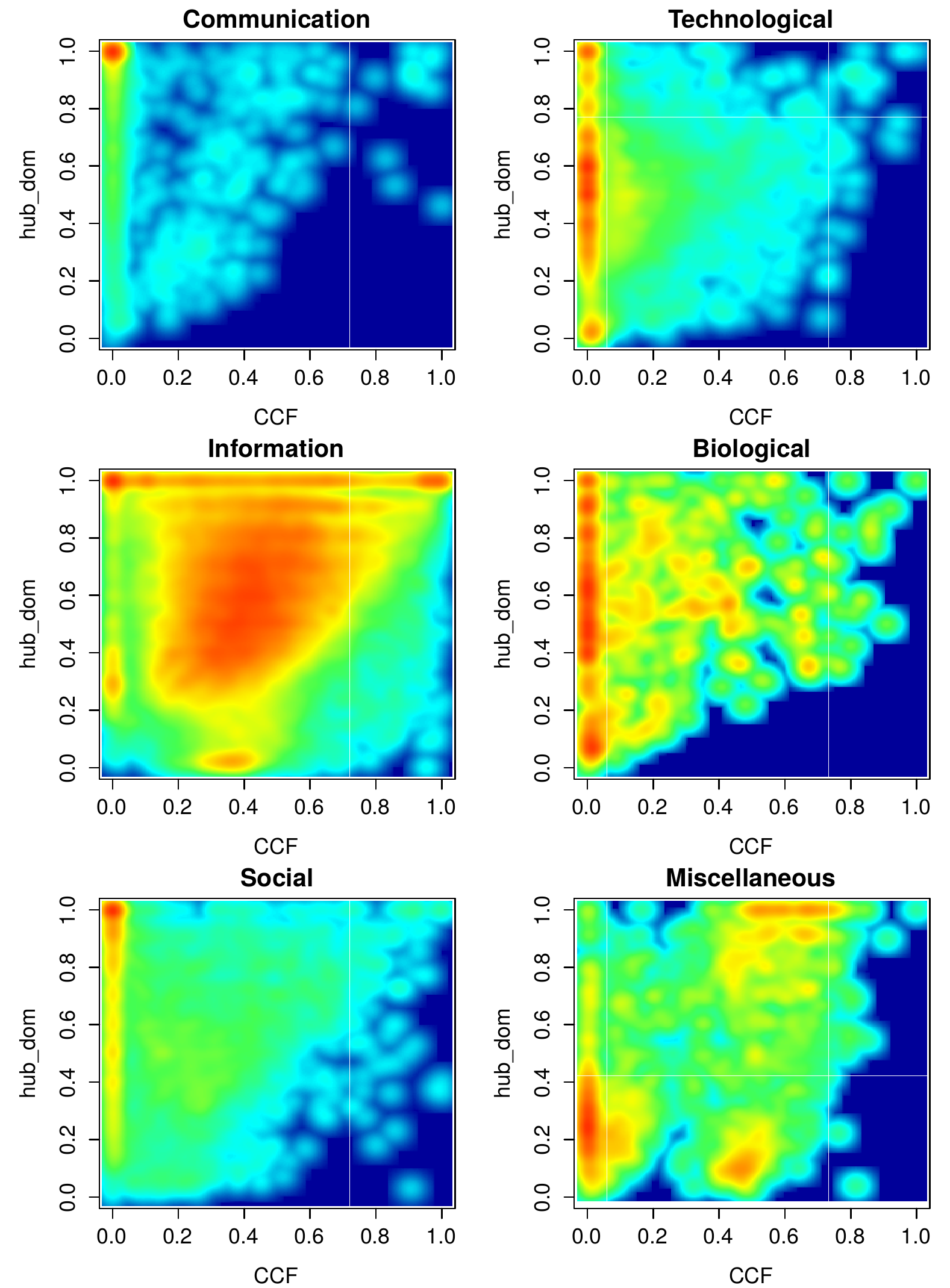}
\caption{Heat maps of distributions of large structural communities detected on different categories of networks are presented on a two dimensional space characterized by transitivity (CCF) and hub dominance (hub\_dom). Only communities of \textbf{more than 10 nodes} are included. From left to right, top to bottom (a) Communication, (b) Technological, (c) Information, (d) Biological, (e) Social, (f) Miscellaneous consists in power networks, ecological networks, artificial networks, etc. }
\label{fig:ccf_hub_large}
\end{figure}

In this section, we show empirical evidences to associate structural communities in real world networks with corresponding topologies determined by the bivariate representation. In order to do that, first \textsf{CCF} and \textsf{hub\_dom} quality scores are calculated over the whole set of communities detected on the network dataset by the presented algorithms. Later, these communities are located in the characterized space in function of their couples of values \textsf{(CCF,hub\_dom)} which represent transitivity and hub dominance respectively. The distribution of communities on this two dimensional space helps to match the most corresponding topologies with each set of communities thanks to the topology characterization presented in the previous section. Since it has been noticed that some structural characteristics might differ between small communities called \textit{micro-communities} and large communities called \textit{macro-communities}~\cite{lancichinetti:2010a}, we proceed to analyze them separately. Figure~\ref{fig:ccf_hub_small} delineates the distributions of small communities of 10 nodes or less in 6 different network groups including communication, technological, information, biological, social and miscellaneous networks as described in Table~\ref{tab:dataset}. The homologous distributions for large communities of more than 10 nodes are depicted in Figure~\ref{fig:ccf_hub_large}. 

At a first sight, it is easy to remark that there is a much higher diversity of structures at the large scale communities than at the small scale communities as the distributions are much more expanded over the space in the former case. It is reasonable since there are much more possibilities how nodes can be connected in a large community than in a small one. Hence large communities' structures are more distinctive and at the same time more complex. Specifically, most of small communities are found around two axis where \textsf{CCF} $=0$ or \textsf{hub\_dom} $=1$, especially at their crosspoint where \textsf{CCF} $=0$ and \textsf{hub\_dom} $=1$. It means star-based and hub dominated structures are very well representative for small communities of every network category. On the other hand, grid structure is totally absent at this size scale, which is quite predictable since it requires a large number of nodes for a grid to be formed. Additionally, the heavy-tail degree distribution recognized in many real world networks make grids less likely to be established.

In information and miscellaneous groups, communities are much more rich in structure comparing to the other categories at both scales. Concretely, besides star-like modules, there are also many clique-like communities and mixture structures since clustering coefficient values in these groups stretch across the whole range. Similarly for hub dominance values which are measured approximately from $0.4$ to $1$ at the small scale and from $0$ to $1$ at the large scale. Although there are some differences in community structure between various network categories, at a small scale, it not very obvious to distinguish them using the proposed representation. We introduce in the following part a detail inspection, especially for large communities, which would reveal essential distinctions between community structure of each network category. The distribution of communities over the profiled map characterizes the mesoscopic structural identity of networks.  

\subsection{Communication networks} 
\begin{figure}
\centering\includegraphics[width=0.98\linewidth]{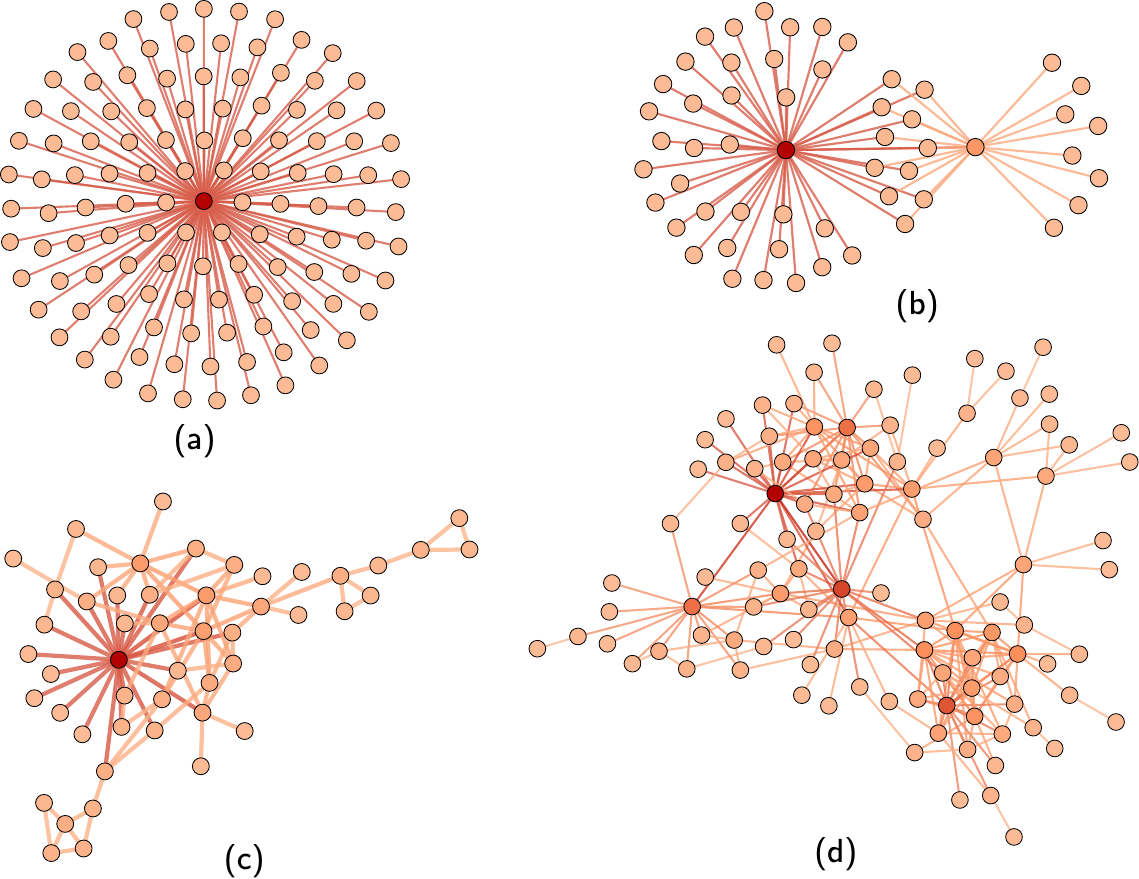}
\caption[Some representative topologies detected in \textit{Communication networks} with their corresponding scores $(CCF, hub\_dom)$. Topologies are ordered from the most famous to the less famous in their network category as shown in Figure~\ref{fig:ccf_hub_large}(a), ~\ref{fig:ccf_hub_small}(a). Hub nodes are darker than peripheral nodes. (a) Email traffic in an European research institution community - $(0,1)$; (b) Wikipedia adminship vote community - $(0.03,0.87)$; (c) Email communication Enron network - $(0.07,0.90)$; (d) Community of email exchange in an university - $(0.28,0.23)$.]{Some representative topologies detected in \textit{Communication networks} with their corresponding scores $(CCF, hub\_dom)$. Topologies are ordered from the most famous to the less famous in their network category as shown in Figure~\ref{fig:ccf_hub_large}(a), ~\ref{fig:ccf_hub_small}(a). Hub nodes are darker than peripheral nodes. (a) Email traffic in an European research institution~\cite{rossi:2015a} community - $(0,1)$; (b) Wikipedia adminship vote~\cite{snapnets} community - $(0.03,0.87)$; (c) Email communication Enron network - $(0.07,0.90)$; (d) Community of email exchange in an university - $(0.28,0.23)$. }
\label{fig:com_topo}
\end{figure}

Communication communities consist in subnetworks of message exchange in social networks, email communications, discussions in forums, etc. From the bivariate distributions of communities shown in Figure~\ref{fig:ccf_hub_large}(a) and~\ref{fig:ccf_hub_small}(a), it can be recognized that structural communities are quite homogeneous in terms of topology in both large and small communities. The majority of them have star-based topologies with very strong hubs which connect to almost every other node in their communities and very few number of clique connections. In other words, communication communities are in general very remarkably high centralized and very low transitive. This property is less clear in large communities than in small communities since the larger a community, the more likely non-hub nodes have chances to create interconnections and possibly establish peripheral hubs. This mechanism also gives rise to a few numbers of multi-hub topologies in large communities. Besides, a small number of hub-absent communities and mesh communities can be discerned. However, they are quite outnumbered by hub structures in this network category. This revelation denotes that exchanges in communication networks often happen around some \textit{central elements} which convey access to their surrounding elements. Figure~\ref{fig:com_topo} illustrates some typical structural community topologies that have been identified in the communication network dataset. Among them, star-like topologies with one dominating hub as shown in Figure~\ref{fig:com_topo}(a),(b) are among the most representative. Besides, there are also communities where hubs are less influential and the presence of a few cliques can be recognized as illustrated in Figure~\ref{fig:com_topo}(c),(d). However, within the list of network categories that has been analyzed in this study, communication communities show a clearest and strongest hub-periphery connection pattern with more than $80\%$ of communities where there are at least 1 node connected to at least $90\%$ of node members in its community and very few periphery-periphery connections. By consequence, communication communities are commonly quite sparse in comparison to other types of networks. Moreover, previous study demonstrated in Figure~\ref{fig:category} helps to infer that communities networks reveal strong \textit{scale-free} property. Consequently, a preferential attachment mechanism with an amplified connection probability to hub nodes would efficiently mimic the structure of real world communication networks. 

\subsection{Technological networks} 
\begin{figure}
\centering\includegraphics[width=0.99\linewidth]{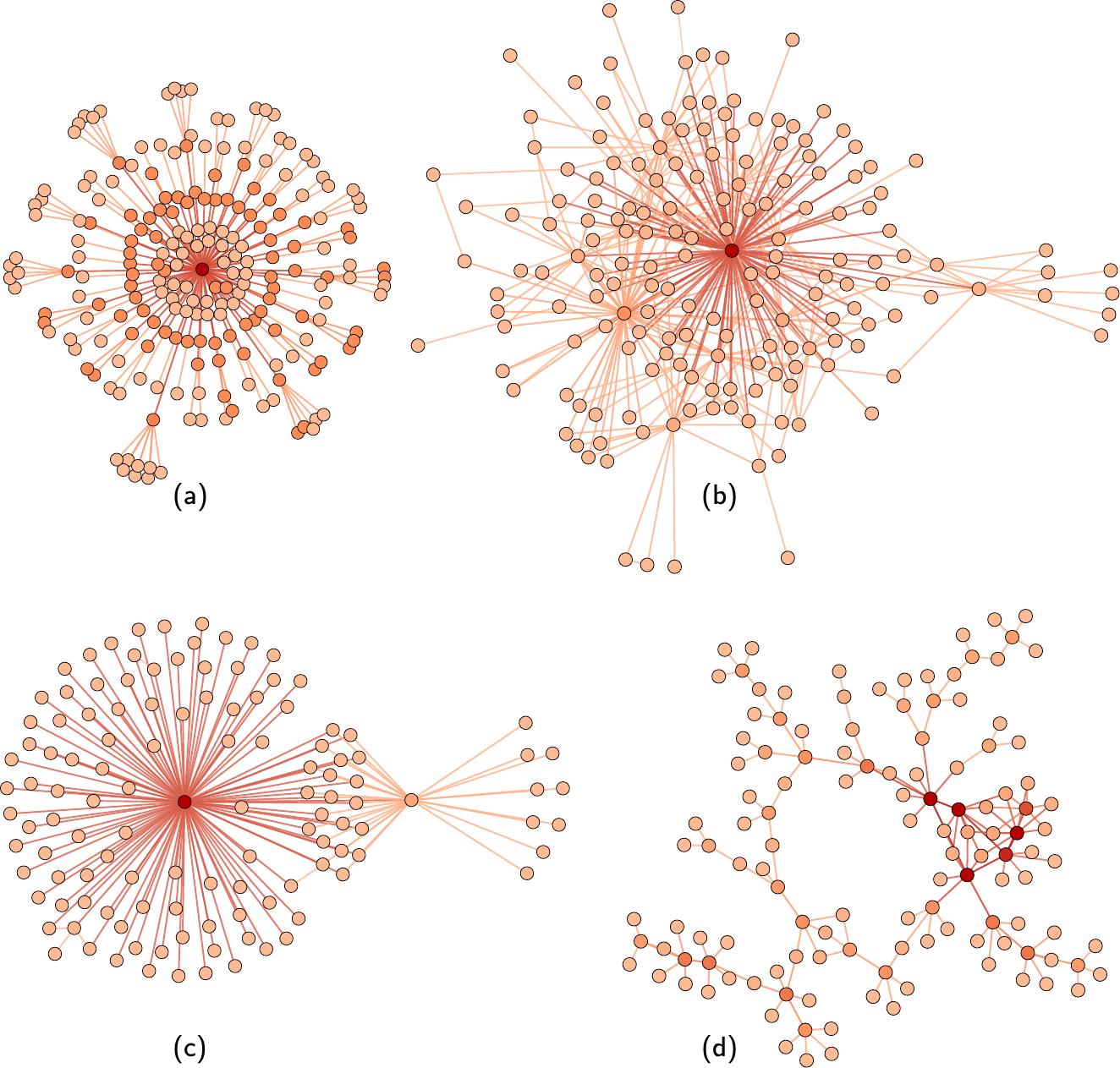}
\caption{Some representative topologies detected in \textit{Technological networks} with their corresponding scores $(CCF, hub\_dom)$. Topologies are ordered from the most famous to the less famous in their network category as shown in Figure~\ref{fig:ccf_hub_small}(b), ~\ref{fig:ccf_hub_large}(b). Hub nodes are darker than peripheral nodes. (a) A community of users of the Pretty-Good-Privacy algorithm for secure information interchange - $(0.01,0.48)$; (b) WHOIS Internet IP community - $(0.07,0.65)$; (c) A community of AS Caida Internet infrastructure recorded in 2007 - $(0.01,0.92)$; (d) A Gnutella peer-to peer network community - $(0.01,0.07)$.}
\label{fig:tech_topo}
\end{figure}

Technological communities include subnetworks in peer-to-peer Gnutella file sharing networks, Internet, highway and airport circulation systems, etc. The most notable similarity between technological communities and communication communities is the high presence of hub-based topologies, especially in small communities as can be seen in Figure~\ref{fig:ccf_hub_small}(b). In large communities, however, technological communities show a quite discernible connection pattern as hubs are less \textit{powerful} in their local as can be interpreted from Figure~\ref{fig:ccf_hub_large}(b). Quantitatively, the majority of hubs in technological networks embrace around $40\%$ to $60\%$ of nodes in their communities. Additionally, the withdraw of super dominating hubs is replaced by the occurrence of more triadic connections in technological communities. It can be explained by the fact that in some infrastructure networks such as highway networks or the Internet, hubs are often constructed to have a controlled influence and are normally compensated by resilient connections or supplement hubs in order to reduce workload, vulnerability or crucial impact caused by their dysfunctionality. Figure~\ref{fig:tech_topo} illustrates some community topologies that have been identified in the technological network dataset. Topologies whose hubs connect to around a half of node members as depicted in Figure~\ref{fig:tech_topo}(a),(b) are among the most representative of networks in this class. There is usually a stratification in the connection pattern as many nodes are connected to a central node by intermediate nodes. This phenomenon can be considered as a presence of hierarchical organization frequently found in technological systems. Besides, there is also a considerable number of star-based structures such as those of communication case and string-based structures as shown in Figure~\ref{fig:tech_topo}(c) and \ref{fig:tech_topo}(d) respectively. In a general view, the scale free property is quite clear although hub attractiveness is relatively reduced comparing to communication networks. A preferentially attachment fitness provided by a model such as Barab{\'a}si-Albert would allow to imitate well technological structural networks. 

\subsection{Information networks}
\begin{figure}
\centering\includegraphics[width=1.03\linewidth]{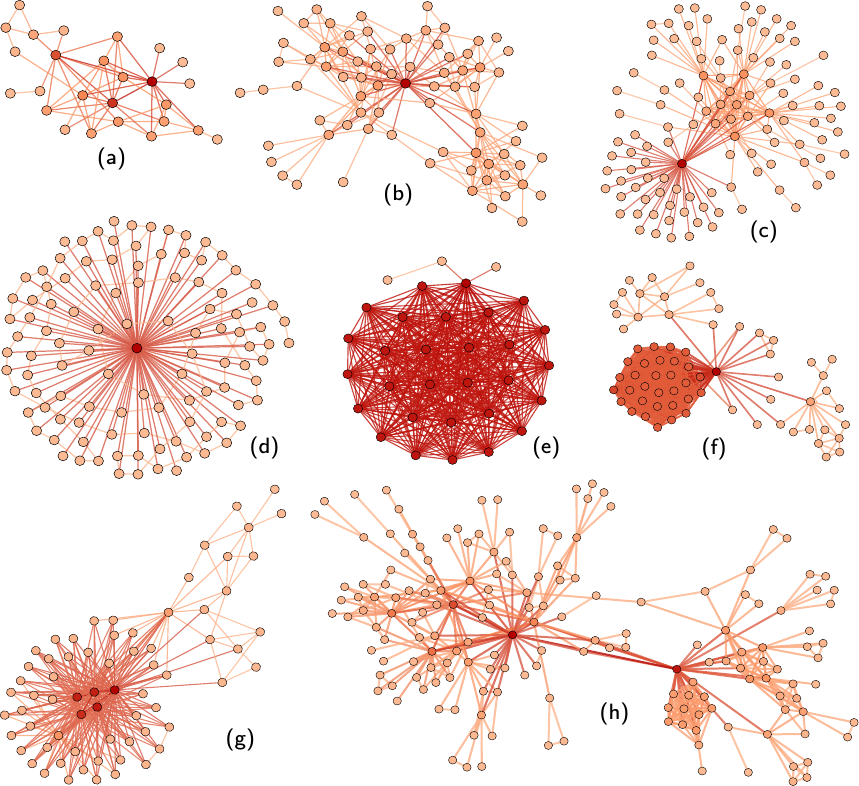}
\caption{Some representative topologies detected in \textit{Information networks} with their corresponding scores $(CCF, hub\_dom)$. Topologies are ordered from the most famous to the less famous in their network category as shown in Figure~\ref{fig:ccf_hub_small}(c), ~\ref{fig:ccf_hub_large}(c). Hub nodes are darker than peripheral nodes. (a,b,g) Amazon recommendation groups of products - $(0.40,0.52), (0.33,0.45)$ and $(0.24,0.76)$ respectively; (c) An educational web system cluster -  $(0.30,0.43)$; (d) A group of Indochina websites recorded in 2004 - $(0.05,0.98)$; (e-f) A community of Arxiv High Energy Physics collaboration - $(0.99,0.97)$ and $(0.95,0.99)$; (h) A collaboration community of Arxiv Condensed Matter network - $(0.44,0.36)$.}
\label{fig:inf_topo}
\end{figure}

Information communities contain subnetworks in citation networks, scientific collaboration networks, research engine networks, recommendation networks, etc. Within the studied networks, information networks exhibit the most diverse topological pattern with the bivariate distribution of communities expanded over a wide range of hub dominance axis and transitivity axis as shown in Figure~\ref{fig:ccf_hub_small}(c),~\ref{fig:ccf_hub_large}(c). Globally, information communities are different from communities of the other network categories by their high transitivity. Such that cliques are very well presented in many information networks as depicted in Figure~\ref{fig:inf_topo}. Many information communities 
can be considered as mixtures of different basic topologies of star-based, string-based, clique-based and grid-based such as the community of collaboration in Arxiv Condensed Matter network shown in Figure~\ref{fig:inf_topo}(h). The presence of hubs in information networks is still high, however they are not anymore the only elements who connect different members of networks. Consequently, information networks are normally much more dense and well connected than other types of networks of the same size scale. This is probably the most representative connectivity feature of information networks. Similar results related to dense and clique structures have been also found by Lancichinetti \textit{et al.}~\cite{lancichinetti:2010a}. Figure~\ref{fig:inf_topo}(a-h) depict some representative communities that have been discovered in some information networks. While the structure in Figure~\ref{fig:inf_topo}(d) resembles a star-based topology with a sequence of periphery-periphery connections; the one in Figure~\ref{fig:inf_topo}(e) of Arxiv High Energy Physics collaboration looks like a complete network with some ill-connected nodes. Figure~\ref{fig:inf_topo}(c,g) demonstrating web and recommendation systems reveal a mixture structure where hubs can be well recognized and clique presence is also remarkable at the same time. The hybrid structure is globally more blended in communities of Figure~\ref{fig:inf_topo}(a,b,h) than the others. The diversity in the structure of information networks can be explained by the way we define this category. In fact, a commercial recommendation system could be very unalike a web citation or a collaboration network, even though they are all considered to be information systems in the network science community. Furthermore, their structures are normally exposed to several complex phenomena that regulate network interactions. Hence, simulating information networks merits more investigation on each concrete case to determine principle mechanism that reflects well the mesoscopic organization.  

\subsection{Biological networks} 
\begin{figure}
\centering\includegraphics[width=1.04\linewidth]{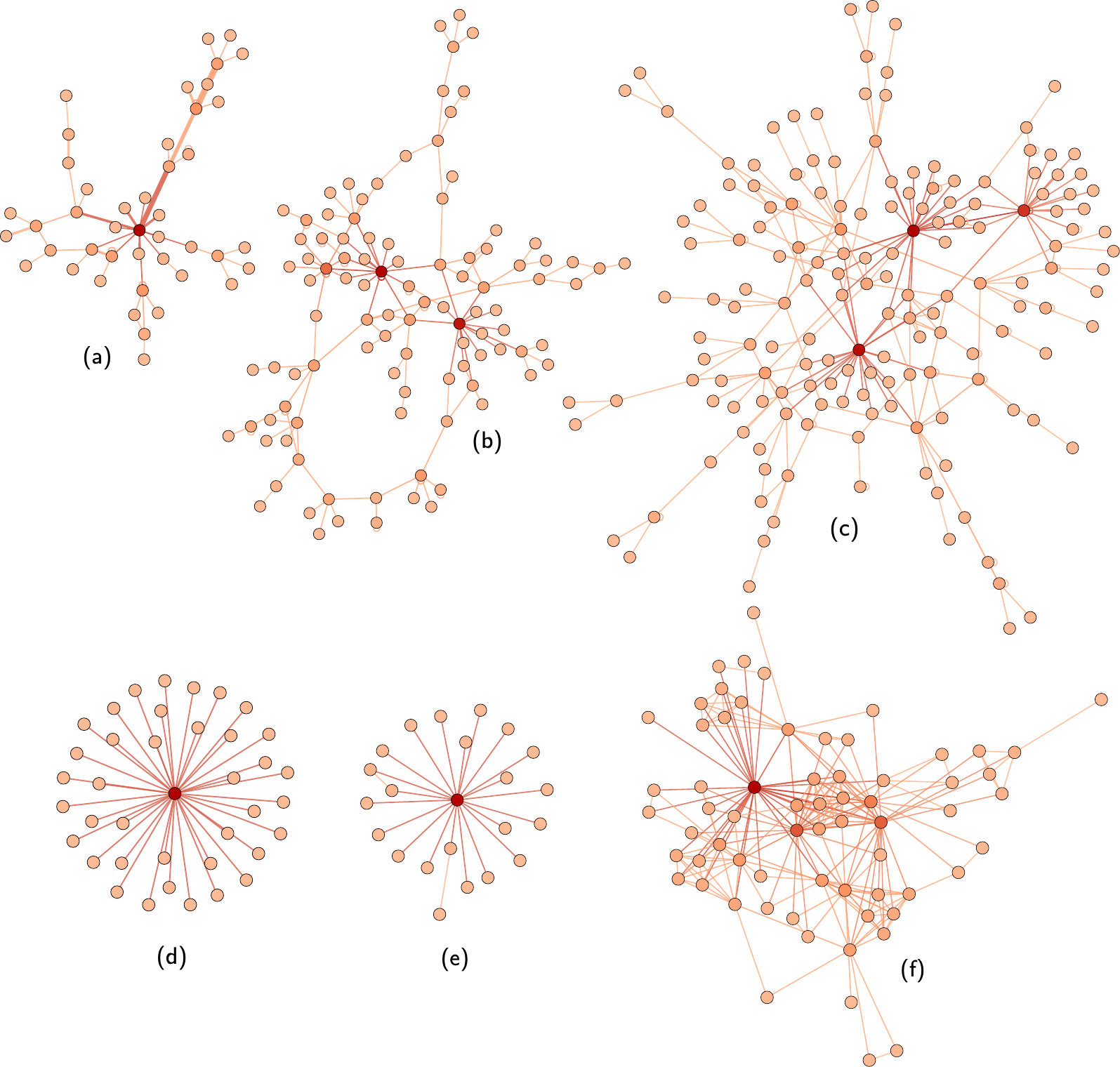}
\caption{Some representative topologies detected in \textit{Biological networks} with their corresponding scores $(CCF, hub\_dom)$. Topologies are ordered from the most famous to the less famous in their network category as shown in Figure~\ref{fig:ccf_hub_small}(d), ~\ref{fig:ccf_hub_large}(d). Hub nodes are darker than peripheral nodes. (a) A circuit of medulla of drosophila fly brain - $(0.06,0.44)$; (b-c) A protein-protein interaction network of yeast - $(0.03, 0.16)$ and $(0.05,0.16)$ respectively; (d-e) protein interactions of drosophila melanogaster $(0,1)$ and $(0.01,0.95)$; (f) A cluster of human disease network $(0.47,0.51)$.}
\label{fig:bio_topo}
\end{figure}

Biological communities comprise subnetworks in brain networks, yeast networks, protein-protein interaction networks, metabolic reaction networks, etc. In some ways, their topologies resemble with technological networks as it can be observed through their distributions in Figure~\ref{fig:ccf_hub_large}(b) and~\ref{fig:ccf_hub_large}(d). The most remarkable discrimination of connection pattern between biological networks with the other ones are their string-based rich structure as can be seen through communities shown in Figure~\ref{fig:bio_topo}(a), (b), (c). The high presence of chains or strings in biological networks has been also found by the other studies using different approaches such as in~\cite{lancichinetti:2010a}, ~\cite{guimera:2007a}. This may be caused by the fact that many biological pathways, which are series of molecular interactions, are included in the analysis and contribute to the high presence of strings. Additionally, many biological networks are only constructed partially due to high complexity in construction time and technical constraints in biochemistry~\cite{newman:2010a}. Therefore, we often observe and analyze small fragments of networks where many connections are missing.

Still, there exist biological networks whose topologies are star-based or hybrid as those of communication networks, technological networks or information networks. However, the hub dominance is globally less important as biological communities are normally small and hubs connect to much less number of their surrounding neighbors. A local observation on biological networks probably discloses random structures in many parts of the networks although hubs are still well widespread. This emergence of random structures could be the most typical characteristic that differs biological networks from the others. Finally, popular properties such as \textit{scale-free}, \textit{small-world} are less significant in biological class than in information or technological class. 

\subsection{Social networks} 
\begin{figure}
\centering\includegraphics[width=0.85\linewidth]{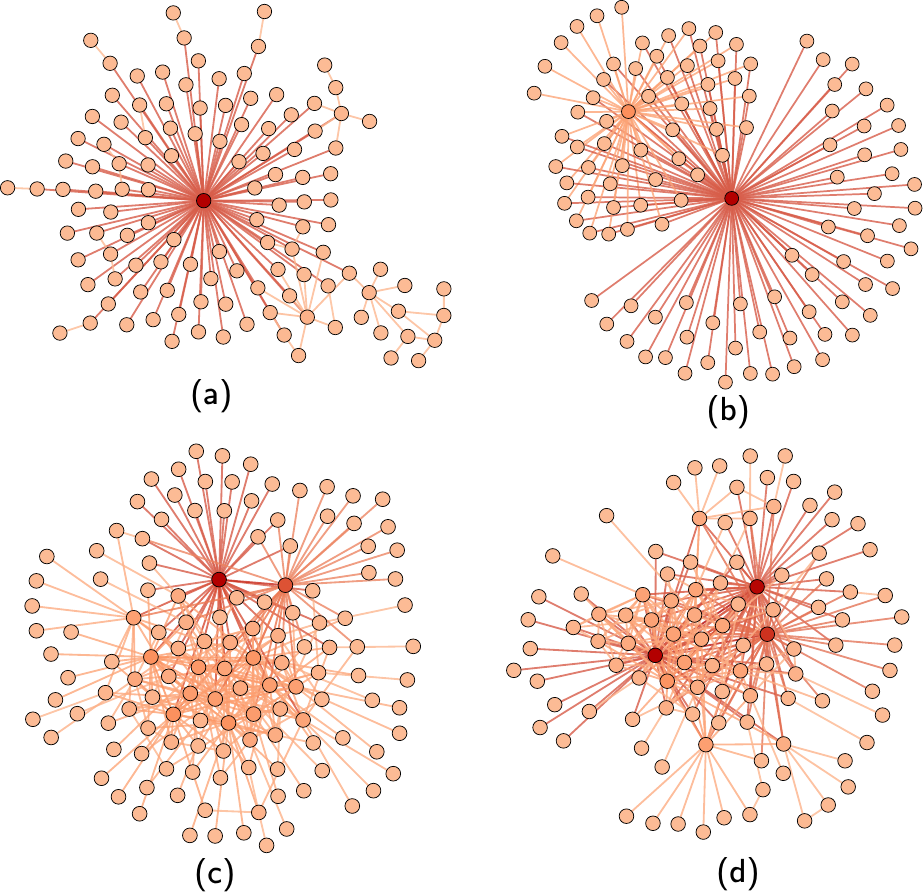}
\caption{Some representative topologies detected in \textit{Social networks} with their corresponding scores $(CCF, hub\_dom)$. Topologies are ordered from the most famous to the less famous in their network category as shown in Figure~\ref{fig:ccf_hub_small}(e), ~\ref{fig:ccf_hub_large}(e). Hub nodes are darker than peripheral nodes. (a) A structural community in Youtube video sharing friendship network - $(0.01,0.81)$; (b) A community in Google Plus network - $(0.02,0.95)$; (c) A political re-tweet network in Twitter - $(0.12,0.60)$ ; (d) A subnetwork of location-based social networking Brightkite - $(0.27,0.51)$.}
\label{fig:soc_topo}
\end{figure}

Social communities involve subnetworks of friendship networks, share or re-tweet networks, followings in Google Plus, Facebook, Twitter, Youtube, etc. Our analysis shows a high similarity in the distribution of large communities in the social networks and communication networks as depicted by Figure~\ref{fig:ccf_hub_large}(a),~Figure~\ref{fig:ccf_hub_large}(e). For small communities, social networks are closer to technological networks and biological networks as shown in Figure~\ref{fig:ccf_hub_small}(b),~\ref{fig:ccf_hub_small}(d),~\ref{fig:ccf_hub_small}(e). A reasonable explanation for the popularity of the star-based topology in social network is that there are many well-known users who are followed or subscribed by a large number of peoples and are becoming mega-connected hubs. Additionally, many samples of social networks that are studied consist of ego networks of celebrities in social media, which makes them intrinsically high centralized around some mega-hub nodes. The only difference with communication communities that has been found in this study is that there are generally more connections between peripheral nodes in social communities. This can be interpreted that friendship or following interactions are generally more frequent than communication interactions. Although different networks of social and communication have been used in this analysis, it makes sense to explain that many users are connected in a social media without or very few communicating interactions in the same channel. For example, two users could be connected on Facebook as friends, but they never exchange any message on the Facebook conversation platform which makes the number of social connections exceeds the number of communications. Figure~\ref{fig:soc_topo} demonstrates some popular topologies of communities in social networks. Note that these topologies are not chosen to argument the differences between various social networks and it is not the objective of this study. They are listed to illustrate some typical and representative structural communities that we discover in the network dataset. Social networks show a clear \textit{scale-free} property as in communication and technological networks, however they are less affected by mega-hubs  and are partially occupied by clique-based structures and many random connections like that of \textit{small-world} phenomenon.

\subsection{Miscellaneous networks} 
\begin{figure}
\centering\includegraphics[width=0.90\linewidth]{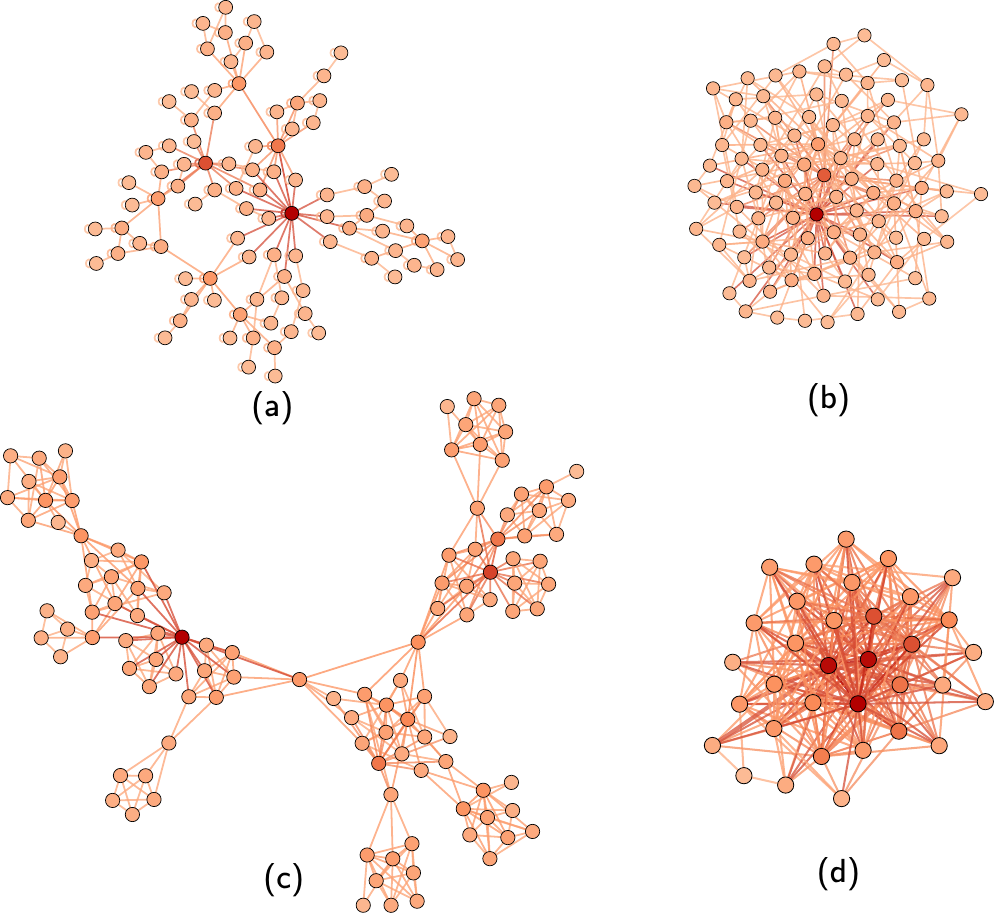}
\caption[Some representative topologies detected in miscellaneous group with their corresponding scores $(CCF, hub\_dom)$. Topologies are ordered from the most famous to the less famous in their network category as shown in Figure~\ref{fig:ccf_hub_small}(f), ~\ref{fig:ccf_hub_large}(f). Hub nodes are darker than peripheral nodes. (a) A cluster of a power network system - $(0.07,0.21)$; (b) A quadratic sieve of a factorization of a 130 bit number - $(0.08,0.39)$; (c) A cluster of a Lancichinetti-Fortunato-Radicchi (LFR) synthetic network - $(0.56,0.18)$; (d) A cluster in an ecological network - $(0.51,0.94)$.]{Some representative topologies detected in miscellaneous group with their corresponding scores $(CCF, hub\_dom)$. Topologies are ordered from the most famous to the less famous in their network category as shown in Figure~\ref{fig:ccf_hub_small}(f), ~\ref{fig:ccf_hub_large}(f). Hub nodes are darker than peripheral nodes. (a) A cluster of a power network system - $(0.07,0.21)$; (b) A quadratic sieve of a factorization of a 130 bit number - $(0.08,0.39)$; (c) A cluster of a Lancichinetti-Fortunato-Radicchi (LFR) synthetic network~\cite{LFR:2008} - $(0.56,0.18)$; (d) A cluster in an ecological network - $(0.51,0.94)$.}
\label{fig:misc_topo}
\end{figure}

Miscellaneous communities cover subnetworks in ecological networks, some power system networks, sport competition networks, synthetic networks, etc. Here, we find many structures, especially in Lancichinetti-Fortunato-Radicchi (LFR) synthetic networks~\cite{LFR:2008}, that are not very popular in the previously studied networks. Specifically, except for information networks, structural communities in the other types of networks are usually very hub-centralized and relatively low in transitivity. On the contrary, in LFR networks, cliques are quite popular and normally aggregated to produce compacted structures as illustrated in Figure~\ref{fig:misc_topo}(c), which makes the communities highly transitive. Additionally, although structures of LFR networks are regulated by many configuration parameters, their hubs generally have less impact in their neighborhoods than those of real world networks such as in social or communication. This is one property that makes a huge difference between LFR benchmarking networks and real world networks. Some other discovered structural communities are illustrated in Figure~\ref{fig:misc_topo}. In a general view, community detection methods identified well compacted sub-graphs in most of the cases.

\section{Discussion and perspectives}
\label{sec:discussion}
In this paper, we provide a novel analysis process to categorize mesoscopic organization of networks into four essential topological groups which show different node organizations. Each representative group is then associated to the corresponding graph generative model that produces a high similarity in connection patterns. Surprisingly, our empirical study uncovers that networks across different categories including communication, technological, information, biological and social networks might have different community structures and can be described by distinguishable characterized topologies. 

The difference of modular topology between networks in various categories could help to construct network profiles or network signatures by domain of study, and hence open a possibility for creating adapted network generative models, network class prediction algorithms, dynamical processes simulation and analysis, etc. Specifically, since networks in each domain reveal some particular modular structures, the mechanisms which are responsible for their creations, evolutions, degradations are also discernible. Hence, different simulation or analysis strategies will generate different impacts on the networks in a predictable way if their structures are well understood. In other words, the network structure profiling assists to achieve suitable network analysis processes and to interpret obtained results without requiring expensive brute force analysis.

In this experiment, we include many state-of-the-art community detection methods whose approaches are quite distinct in order to exploit different facets of modular structure that could be detected on the networks. Nevertheless, the impact of these methods on the revealed structures merits to be examined in more details. An interesting perspective could be the relation between different topologies and different mechanisms that are responsible for identifying or transforming networks/sub-networks from one type to another. Such that, the understanding these mechanisms could help us to explain the effects of community detection algorithms on the partitioning of networks and also how different dynamical processes influence evolving networks.


\begin{thebibliography}{41}%
\makeatletter
\providecommand \@ifxundefined [1]{%
 \@ifx{#1\undefined}
}%
\providecommand \@ifnum [1]{%
 \ifnum #1\expandafter \@firstoftwo
 \else \expandafter \@secondoftwo
 \fi
}%
\providecommand \@ifx [1]{%
 \ifx #1\expandafter \@firstoftwo
 \else \expandafter \@secondoftwo
 \fi
}%
\providecommand \natexlab [1]{#1}%
\providecommand \enquote  [1]{``#1''}%
\providecommand \bibnamefont  [1]{#1}%
\providecommand \bibfnamefont [1]{#1}%
\providecommand \citenamefont [1]{#1}%
\providecommand \href@noop [0]{\@secondoftwo}%
\providecommand \href [0]{\begingroup \@sanitize@url \@href}%
\providecommand \@href[1]{\@@startlink{#1}\@@href}%
\providecommand \@@href[1]{\endgroup#1\@@endlink}%
\providecommand \@sanitize@url [0]{\catcode `\\12\catcode `\$12\catcode
  `\&12\catcode `\#12\catcode `\^12\catcode `\_12\catcode `\%12\relax}%
\providecommand \@@startlink[1]{}%
\providecommand \@@endlink[0]{}%
\providecommand \url  [0]{\begingroup\@sanitize@url \@url }%
\providecommand \@url [1]{\endgroup\@href {#1}{\urlprefix }}%
\providecommand \urlprefix  [0]{URL }%
\providecommand \Eprint [0]{\href }%
\providecommand \doibase [0]{http://dx.doi.org/}%
\providecommand \selectlanguage [0]{\@gobble}%
\providecommand \bibinfo  [0]{\@secondoftwo}%
\providecommand \bibfield  [0]{\@secondoftwo}%
\providecommand \translation [1]{[#1]}%
\providecommand \BibitemOpen [0]{}%
\providecommand \bibitemStop [0]{}%
\providecommand \bibitemNoStop [0]{.\EOS\space}%
\providecommand \EOS [0]{\spacefactor3000\relax}%
\providecommand \BibitemShut  [1]{\csname bibitem#1\endcsname}%
\let\auto@bib@innerbib\@empty
%</preamble>
\bibitem [{\citenamefont {Leskovec}\ \emph
  {et~al.}(2008{\natexlab{a}})\citenamefont {Leskovec}, \citenamefont
  {Backstrom}, \citenamefont {Kumar},\ and\ \citenamefont
  {Tomkins}}]{leskovec:2008b}%
  \BibitemOpen
  \bibfield  {author} {\bibinfo {author} {\bibfnamefont {J.}~\bibnamefont
  {Leskovec}}, \bibinfo {author} {\bibfnamefont {L.}~\bibnamefont {Backstrom}},
  \bibinfo {author} {\bibfnamefont {R.}~\bibnamefont {Kumar}}, \ and\ \bibinfo
  {author} {\bibfnamefont {A.}~\bibnamefont {Tomkins}},\ }in\ \href {\doibase
  10.1145/1401890.1401948} {\emph {\bibinfo {booktitle} {Proceedings of the
  14th ACM SIGKDD International Conference on Knowledge Discovery and Data
  Mining}}},\ \bibinfo {series and number} {KDD '08}\ (\bibinfo  {publisher}
  {ACM},\ \bibinfo {address} {New York, NY, USA},\ \bibinfo {year} {2008})\
  pp.\ \bibinfo {pages} {462--470}\BibitemShut {NoStop}%
\bibitem [{\citenamefont {{Yang}}\ and\ \citenamefont
  {{Leskovec}}(2015)}]{yang:2015a}%
  \BibitemOpen
  \bibfield  {author} {\bibinfo {author} {\bibfnamefont {J.}~\bibnamefont
  {{Yang}}}\ and\ \bibinfo {author} {\bibfnamefont {J.}~\bibnamefont
  {{Leskovec}}},\ }\href@noop {} {\bibfield  {journal} {\bibinfo  {journal}
  {Knowledge and Information Systems}\ }\textbf {\bibinfo {volume} {42}}
  (\bibinfo {year} {2015})},\ \Eprint {http://arxiv.org/abs/1205.6233}
  {1205.6233} \BibitemShut {NoStop}%
\bibitem [{\citenamefont {Eric}\ and\ \citenamefont {Adam}(2003)}]{eric:2003a}%
  \BibitemOpen
  \bibfield  {author} {\bibinfo {author} {\bibfnamefont {A.}~\bibnamefont
  {Eric}}\ and\ \bibinfo {author} {\bibfnamefont {P.~A.}\ \bibnamefont
  {Adam}},\ }\href {\doibase https://doi.org/10.1016/S0959-440X(03)00031-9}
  {\bibfield  {journal} {\bibinfo  {journal} {{Current Opinion in Structural
  Biology}}\ }\textbf {\bibinfo {volume} {13}},\ \bibinfo {pages} {193 }
  (\bibinfo {year} {2003})}\BibitemShut {NoStop}%
\bibitem [{\citenamefont {R{\'e}ka}(2005)}]{albert:2005a}%
  \BibitemOpen
  \bibfield  {author} {\bibinfo {author} {\bibfnamefont {A.}~\bibnamefont
  {R{\'e}ka}},\ }\href {\doibase 10.1242/jcs.02714} {\bibfield  {journal}
  {\bibinfo  {journal} {Journal of Cell Science}\ }\textbf {\bibinfo {volume}
  {118}},\ \bibinfo {pages} {4947} (\bibinfo {year} {2005})}\BibitemShut
  {NoStop}%
\bibitem [{\citenamefont {Xiaowei}\ \emph {et~al.}(2007)\citenamefont
  {Xiaowei}, \citenamefont {Mark},\ and\ \citenamefont {Michael}}]{zhu:2007a}%
  \BibitemOpen
  \bibfield  {author} {\bibinfo {author} {\bibfnamefont {Z.}~\bibnamefont
  {Xiaowei}}, \bibinfo {author} {\bibfnamefont {G.}~\bibnamefont {Mark}}, \
  and\ \bibinfo {author} {\bibfnamefont {S.}~\bibnamefont {Michael}},\
  }\href@noop {} {\bibfield  {journal} {\bibinfo  {journal} {Genes and
  Development}\ }\textbf {\bibinfo {volume} {21(9)}},\ \bibinfo {pages} {1010}
  (\bibinfo {year} {2007})}\BibitemShut {NoStop}%
\bibitem [{\citenamefont {Faloutsos}\ \emph {et~al.}(1999)\citenamefont
  {Faloutsos}, \citenamefont {Faloutsos},\ and\ \citenamefont
  {Faloutsos}}]{faloutsos:1999a}%
  \BibitemOpen
  \bibfield  {author} {\bibinfo {author} {\bibfnamefont {M.}~\bibnamefont
  {Faloutsos}}, \bibinfo {author} {\bibfnamefont {P.}~\bibnamefont
  {Faloutsos}}, \ and\ \bibinfo {author} {\bibfnamefont {C.}~\bibnamefont
  {Faloutsos}},\ }\href {\doibase 10.1145/316194.316229} {\bibfield  {journal}
  {\bibinfo  {journal} {{SIGCOMM Comput. Commun. Rev.}}\ }\textbf {\bibinfo
  {volume} {29}},\ \bibinfo {pages} {251} (\bibinfo {year} {1999})}\BibitemShut
  {NoStop}%
\bibitem [{\citenamefont {V\'azquez}\ \emph {et~al.}(2002)\citenamefont
  {V\'azquez}, \citenamefont {Pastor-Satorras},\ and\ \citenamefont
  {Vespignani}}]{vazquez:2002a}%
  \BibitemOpen
  \bibfield  {author} {\bibinfo {author} {\bibfnamefont {A.}~\bibnamefont
  {V\'azquez}}, \bibinfo {author} {\bibfnamefont {R.}~\bibnamefont
  {Pastor-Satorras}}, \ and\ \bibinfo {author} {\bibfnamefont {A.}~\bibnamefont
  {Vespignani}},\ }\href {\doibase 10.1103/PhysRevE.65.066130} {\bibfield
  {journal} {\bibinfo  {journal} {Phys. Rev. E}\ }\textbf {\bibinfo {volume}
  {65}},\ \bibinfo {pages} {066130} (\bibinfo {year} {2002})}\BibitemShut
  {NoStop}%
\bibitem [{\citenamefont {{Raghavan}}\ \emph {et~al.}(2007)\citenamefont
  {{Raghavan}}, \citenamefont {{Albert}},\ and\ \citenamefont
  {{Kumara}}}]{raghavan:2007a}%
  \BibitemOpen
  \bibfield  {author} {\bibinfo {author} {\bibfnamefont {U.~N.}\ \bibnamefont
  {{Raghavan}}}, \bibinfo {author} {\bibfnamefont {R.}~\bibnamefont
  {{Albert}}}, \ and\ \bibinfo {author} {\bibfnamefont {S.}~\bibnamefont
  {{Kumara}}},\ }\href {\doibase 10.1103/PhysRevE.76.036106} {\bibfield
  {journal} {\bibinfo  {journal} {Phys. Rev. E}\ }\textbf {\bibinfo {volume}
  {76}},\ \bibinfo {eid} {036106} (\bibinfo {year} {2007})},\ \Eprint
  {http://arxiv.org/abs/0709.2938} {0709.2938} \BibitemShut {NoStop}%
\bibitem [{\citenamefont {Girvan}\ and\ \citenamefont
  {Newman}(2002)}]{girvan:2002a}%
  \BibitemOpen
  \bibfield  {author} {\bibinfo {author} {\bibfnamefont {M.}~\bibnamefont
  {Girvan}}\ and\ \bibinfo {author} {\bibfnamefont {M.~E.~J.}\ \bibnamefont
  {Newman}},\ }\href@noop {} {\bibfield  {journal} {\bibinfo  {journal}
  {Proceedings of the National Academy of Sciences}\ }\textbf {\bibinfo
  {volume} {99}},\ \bibinfo {pages} {7821} (\bibinfo {year}
  {2002})}\BibitemShut {NoStop}%
\bibitem [{\citenamefont {{Newman}}\ and\ \citenamefont
  {{Girvan}}(2004)}]{newman:2004a}%
  \BibitemOpen
  \bibfield  {author} {\bibinfo {author} {\bibfnamefont {M.~E.~J.}\
  \bibnamefont {{Newman}}}\ and\ \bibinfo {author} {\bibfnamefont
  {M.}~\bibnamefont {{Girvan}}},\ }\href {\doibase 10.1103/PhysRevE.69.026113}
  {\bibfield  {journal} {\bibinfo  {journal} {{Phys. Rev. E}}\ }\textbf
  {\bibinfo {volume} {69}},\ \bibinfo {eid} {026113} (\bibinfo {year}
  {2004})}\BibitemShut {NoStop}%
\bibitem [{\citenamefont {Rosvall}\ and\ \citenamefont
  {Bergstrom}(2008)}]{rosvall:2008a}%
  \BibitemOpen
  \bibfield  {author} {\bibinfo {author} {\bibfnamefont {M.}~\bibnamefont
  {Rosvall}}\ and\ \bibinfo {author} {\bibfnamefont {C.~T.}\ \bibnamefont
  {Bergstrom}},\ }\href {\doibase 10.1073/pnas.0706851105} {\bibfield
  {journal} {\bibinfo  {journal} {Proceedings of the National Academy of
  Sciences}\ }\textbf {\bibinfo {volume} {105}},\ \bibinfo {pages} {1118}
  (\bibinfo {year} {2008})}\BibitemShut {NoStop}%
\bibitem [{\citenamefont {{Leskovec}}\ \emph {et~al.}(2010)\citenamefont
  {{Leskovec}}, \citenamefont {{Lang}},\ and\ \citenamefont
  {{Mahoney}}}]{leskovec:2010a}%
  \BibitemOpen
  \bibfield  {author} {\bibinfo {author} {\bibfnamefont {J.}~\bibnamefont
  {{Leskovec}}}, \bibinfo {author} {\bibfnamefont {K.~J.}\ \bibnamefont
  {{Lang}}}, \ and\ \bibinfo {author} {\bibfnamefont {M.~W.}\ \bibnamefont
  {{Mahoney}}},\ }\href@noop {} {\bibfield  {journal} {\bibinfo  {journal} {ACM
  WWW International Conference on World Wide Web}\ } (\bibinfo {year}
  {2010})},\ \Eprint {http://arxiv.org/abs/1004.3539} {1004.3539} \BibitemShut
  {NoStop}%
\bibitem [{\citenamefont {{Fortunato}}(2010)}]{fortunato:2010a}%
  \BibitemOpen
  \bibfield  {author} {\bibinfo {author} {\bibfnamefont {S.}~\bibnamefont
  {{Fortunato}}},\ }\href {\doibase 10.1016/j.physrep.2009.11.002} {\bibfield
  {journal} {\bibinfo  {journal} {Physics Reports}\ }\textbf {\bibinfo {volume}
  {486}},\ \bibinfo {pages} {75} (\bibinfo {year} {2010})},\ \Eprint
  {http://arxiv.org/abs/0906.0612} {0906.0612} \BibitemShut {NoStop}%
\bibitem [{\citenamefont {{Coscia}}\ \emph {et~al.}(2012)\citenamefont
  {{Coscia}}, \citenamefont {{Giannotti}},\ and\ \citenamefont
  {{Pedreschi}}}]{coscia:2012a}%
  \BibitemOpen
  \bibfield  {author} {\bibinfo {author} {\bibfnamefont {M.}~\bibnamefont
  {{Coscia}}}, \bibinfo {author} {\bibfnamefont {F.}~\bibnamefont
  {{Giannotti}}}, \ and\ \bibinfo {author} {\bibfnamefont {D.}~\bibnamefont
  {{Pedreschi}}},\ }\href@noop {} {\bibfield  {journal} {\bibinfo  {journal}
  {Statistical Analysis and Data Mining}\ } (\bibinfo {year} {2012})},\ \Eprint
  {http://arxiv.org/abs/1206.3552} {arXiv:1206.3552} \BibitemShut {NoStop}%
\bibitem [{\citenamefont {{Rosvall}}\ \emph {et~al.}(2017)\citenamefont
  {{Rosvall}}, \citenamefont {{Delvenne}}, \citenamefont {{Schaub}},\ and\
  \citenamefont {{Lambiotte}}}]{rosvall:2017a}%
  \BibitemOpen
  \bibfield  {author} {\bibinfo {author} {\bibfnamefont {M.}~\bibnamefont
  {{Rosvall}}}, \bibinfo {author} {\bibfnamefont {J.-C.}\ \bibnamefont
  {{Delvenne}}}, \bibinfo {author} {\bibfnamefont {M.~T.}\ \bibnamefont
  {{Schaub}}}, \ and\ \bibinfo {author} {\bibfnamefont {R.}~\bibnamefont
  {{Lambiotte}}},\ }\href@noop {} {\bibfield  {journal} {\bibinfo  {journal}
  {{ArXiv e-prints}}\ } (\bibinfo {year} {2017})},\ \Eprint
  {http://arxiv.org/abs/1712.06468} {1712.06468} \BibitemShut {NoStop}%
\bibitem [{\citenamefont {{Fortunato}}\ and\ \citenamefont
  {{Hric}}(2016)}]{fortunato:2016a}%
  \BibitemOpen
  \bibfield  {author} {\bibinfo {author} {\bibfnamefont {S.}~\bibnamefont
  {{Fortunato}}}\ and\ \bibinfo {author} {\bibfnamefont {D.}~\bibnamefont
  {{Hric}}},\ }\href {\doibase 10.1016/j.physrep.2016.09.002} {\bibfield
  {journal} {\bibinfo  {journal} {{Physics Reports}}\ }\textbf {\bibinfo
  {volume} {659}},\ \bibinfo {pages} {1} (\bibinfo {year} {2016})},\ \Eprint
  {http://arxiv.org/abs/1608.00163} {1608.00163} \BibitemShut {NoStop}%
\bibitem [{\citenamefont {{Lancichinetti}}\ \emph {et~al.}(2010)\citenamefont
  {{Lancichinetti}}, \citenamefont {{Kivel{\"a}}}, \citenamefont
  {{Saram{\"a}ki}},\ and\ \citenamefont {{Fortunato}}}]{lancichinetti:2010a}%
  \BibitemOpen
  \bibfield  {author} {\bibinfo {author} {\bibfnamefont {A.}~\bibnamefont
  {{Lancichinetti}}}, \bibinfo {author} {\bibfnamefont {M.}~\bibnamefont
  {{Kivel{\"a}}}}, \bibinfo {author} {\bibfnamefont {J.}~\bibnamefont
  {{Saram{\"a}ki}}}, \ and\ \bibinfo {author} {\bibfnamefont {S.}~\bibnamefont
  {{Fortunato}}},\ }\href {\doibase 10.1371/journal.pone.0011976} {\bibfield
  {journal} {\bibinfo  {journal} {PLoS ONE}\ }\textbf {\bibinfo {volume} {5}},\
  \bibinfo {pages} {e11976} (\bibinfo {year} {2010})},\ \Eprint
  {http://arxiv.org/abs/1005.4376} {1005.4376} \BibitemShut {NoStop}%
\bibitem [{\citenamefont {{Guimera}}\ \emph {et~al.}(2007)\citenamefont
  {{Guimera}}, \citenamefont {{Sales-Pardo}},\ and\ \citenamefont
  {{Amaral}}}]{guimera:2007a}%
  \BibitemOpen
  \bibfield  {author} {\bibinfo {author} {\bibfnamefont {R.}~\bibnamefont
  {{Guimera}}}, \bibinfo {author} {\bibfnamefont {M.}~\bibnamefont
  {{Sales-Pardo}}}, \ and\ \bibinfo {author} {\bibfnamefont {L.~A.~N.}\
  \bibnamefont {{Amaral}}},\ }\href@noop {} {\bibfield  {journal} {\bibinfo
  {journal} {ArXiv Physics e-prints}\ } (\bibinfo {year} {2007})},\ \Eprint
  {http://arxiv.org/abs/physics/0701149} {physics/0701149} \BibitemShut
  {NoStop}%
\bibitem [{\citenamefont {{Guimer{\`a}}}\ \emph {et~al.}(2004)\citenamefont
  {{Guimer{\`a}}}, \citenamefont {{Sales-Pardo}},\ and\ \citenamefont
  {{Amaral}}}]{guimera:2004a}%
  \BibitemOpen
  \bibfield  {author} {\bibinfo {author} {\bibfnamefont {R.}~\bibnamefont
  {{Guimer{\`a}}}}, \bibinfo {author} {\bibfnamefont {M.}~\bibnamefont
  {{Sales-Pardo}}}, \ and\ \bibinfo {author} {\bibfnamefont {L.~A.~N.}\
  \bibnamefont {{Amaral}}},\ }\href {\doibase 10.1103/PhysRevE.70.025101}
  {\bibfield  {journal} {\bibinfo  {journal} {{Phys. Rev. E}}\ }\textbf
  {\bibinfo {volume} {70}},\ \bibinfo {pages} {025101} (\bibinfo {year}
  {2004})},\ \Eprint {http://arxiv.org/abs/cond-mat/0403660} {cond-mat/0403660}
  \BibitemShut {NoStop}%
\bibitem [{\citenamefont {Leskovec}\ \emph
  {et~al.}(2008{\natexlab{b}})\citenamefont {Leskovec}, \citenamefont {Lang},
  \citenamefont {Dasgupta},\ and\ \citenamefont {Mahoney}}]{leskovec:2008a}%
  \BibitemOpen
  \bibfield  {author} {\bibinfo {author} {\bibfnamefont {J.}~\bibnamefont
  {Leskovec}}, \bibinfo {author} {\bibfnamefont {J.}~\bibnamefont {Lang},
  \bibfnamefont {Kevin}}, \bibinfo {author} {\bibfnamefont {A.}~\bibnamefont
  {Dasgupta}}, \ and\ \bibinfo {author} {\bibfnamefont {M.~W.}\ \bibnamefont
  {Mahoney}},\ }in\ \href {\doibase 10.1145/1367497.1367591} {\emph {\bibinfo
  {booktitle} {Proceedings of the 17th International Conference on World Wide
  Web}}},\ \bibinfo {series and number} {WWW '08}\ (\bibinfo  {publisher}
  {ACM},\ \bibinfo {address} {New York, NY, USA},\ \bibinfo {year} {2008})\
  pp.\ \bibinfo {pages} {695--704}\BibitemShut {NoStop}%
\bibitem [{\citenamefont {{Labatut}}\ and\ \citenamefont {{Keziban
  Orman}}(2017)}]{labatut:2017a}%
  \BibitemOpen
  \bibfield  {author} {\bibinfo {author} {\bibfnamefont {V.}~\bibnamefont
  {{Labatut}}}\ and\ \bibinfo {author} {\bibfnamefont {G.}~\bibnamefont
  {{Keziban Orman}}},\ }\href@noop {} {\bibfield  {journal} {\bibinfo
  {journal} {Encyclopedia of Social Network Analysis and Mining - 2nd Edition}\
  } (\bibinfo {year} {2017})}\BibitemShut {NoStop}%
\bibitem [{\citenamefont {Barrat}\ \emph {et~al.}(2004)\citenamefont {Barrat},
  \citenamefont {Barth{\'e}lemy}, \citenamefont {Pastor-Satorras},\ and\
  \citenamefont {Vespignani}}]{barrat:2004a}%
  \BibitemOpen
  \bibfield  {author} {\bibinfo {author} {\bibfnamefont {A.}~\bibnamefont
  {Barrat}}, \bibinfo {author} {\bibfnamefont {M.}~\bibnamefont
  {Barth{\'e}lemy}}, \bibinfo {author} {\bibfnamefont {R.}~\bibnamefont
  {Pastor-Satorras}}, \ and\ \bibinfo {author} {\bibfnamefont {A.}~\bibnamefont
  {Vespignani}},\ }\href {\doibase 10.1073/pnas.0400087101} {\bibfield
  {journal} {\bibinfo  {journal} {Proceedings of the National Academy of
  Sciences}\ }\textbf {\bibinfo {volume} {101}},\ \bibinfo {pages} {3747}
  (\bibinfo {year} {2004})}\BibitemShut {NoStop}%
\bibitem [{\citenamefont {{Blondel}}\ \emph {et~al.}(2008)\citenamefont
  {{Blondel}}, \citenamefont {{Guillaume}}, \citenamefont {{Lambiotte}},\ and\
  \citenamefont {{Lefebvre}}}]{blondel:2008a}%
  \BibitemOpen
  \bibfield  {author} {\bibinfo {author} {\bibfnamefont {V.~D.}\ \bibnamefont
  {{Blondel}}}, \bibinfo {author} {\bibfnamefont {J.-L.}\ \bibnamefont
  {{Guillaume}}}, \bibinfo {author} {\bibfnamefont {R.}~\bibnamefont
  {{Lambiotte}}}, \ and\ \bibinfo {author} {\bibfnamefont {E.}~\bibnamefont
  {{Lefebvre}}},\ }\href@noop {} {\bibfield  {journal} {\bibinfo  {journal}
  {Journal of Statistical Mechanics: Theory and Experiment}\ }\textbf {\bibinfo
  {volume} {10}},\ \bibinfo {pages} {10008} (\bibinfo {year} {2008})},\ \Eprint
  {http://arxiv.org/abs/0803.0476} {arXiv:0803.0476 [physics.soc-ph]}
  \BibitemShut {NoStop}%
\bibitem [{\citenamefont {Clauset}\ \emph {et~al.}(2004)\citenamefont
  {Clauset}, \citenamefont {Newman},\ and\ \citenamefont
  {Moore}}]{aaron:2004a}%
  \BibitemOpen
  \bibfield  {author} {\bibinfo {author} {\bibfnamefont {A.}~\bibnamefont
  {Clauset}}, \bibinfo {author} {\bibfnamefont {M.~E.~J.}\ \bibnamefont
  {Newman}}, \ and\ \bibinfo {author} {\bibfnamefont {C.}~\bibnamefont
  {Moore}},\ }\href@noop {} {\bibfield  {journal} {\bibinfo  {journal} {{Phys.
  Rev. E}}\ }\textbf {\bibinfo {volume} {70}},\ \bibinfo {pages} {066111}
  (\bibinfo {year} {2004})}\BibitemShut {NoStop}%
\bibitem [{\citenamefont {{Pons}}\ and\ \citenamefont
  {{Latapy}}(2005)}]{pons:2005a}%
  \BibitemOpen
  \bibfield  {author} {\bibinfo {author} {\bibfnamefont {P.}~\bibnamefont
  {{Pons}}}\ and\ \bibinfo {author} {\bibfnamefont {M.}~\bibnamefont
  {{Latapy}}},\ }\href@noop {} {\bibfield  {journal} {\bibinfo  {journal}
  {Computer and Information Sciences - ISCIS 2005}\ } (\bibinfo {year}
  {2005})}\BibitemShut {NoStop}%
\bibitem [{\citenamefont {{Reichardt}}\ and\ \citenamefont
  {{Bornholdt}}(2006)}]{reichardt:2006a}%
  \BibitemOpen
  \bibfield  {author} {\bibinfo {author} {\bibfnamefont {J.}~\bibnamefont
  {{Reichardt}}}\ and\ \bibinfo {author} {\bibfnamefont {S.}~\bibnamefont
  {{Bornholdt}}},\ }\href {\doibase 10.1103/PhysRevE.74.016110} {\bibfield
  {journal} {\bibinfo  {journal} {{Phys. Rev. E}}\ }\textbf {\bibinfo {volume}
  {74}},\ \bibinfo {eid} {016110} (\bibinfo {year} {2006})},\ \Eprint
  {http://arxiv.org/abs/cond-mat/0603718} {cond-mat/0603718} \BibitemShut
  {NoStop}%
\bibitem [{\citenamefont {{Newman}}(2006)}]{newman:2006a}%
  \BibitemOpen
  \bibfield  {author} {\bibinfo {author} {\bibfnamefont {M.~E.~J.}\
  \bibnamefont {{Newman}}},\ }\href {\doibase 10.1103/PhysRevE.74.036104}
  {\bibfield  {journal} {\bibinfo  {journal} {{Phys. Rev. E}}\ }\textbf
  {\bibinfo {volume} {74}},\ \bibinfo {eid} {036104} (\bibinfo {year}
  {2006})},\ \Eprint {http://arxiv.org/abs/physics/0605087} {physics/0605087}
  \BibitemShut {NoStop}%
\bibitem [{\citenamefont {{Orman}}\ \emph {et~al.}(2012)\citenamefont
  {{Orman}}, \citenamefont {{Labatut}},\ and\ \citenamefont
  {{Cherifi}}}]{orman:2012a}%
  \BibitemOpen
  \bibfield  {author} {\bibinfo {author} {\bibfnamefont {G.}~\bibnamefont
  {{Orman}}}, \bibinfo {author} {\bibfnamefont {V.}~\bibnamefont {{Labatut}}},
  \ and\ \bibinfo {author} {\bibfnamefont {H.}~\bibnamefont {{Cherifi}}},\
  }\href@noop {} {\bibfield  {journal} {\bibinfo  {journal} {ArXiv e-prints}\ }
  (\bibinfo {year} {2012})},\ \Eprint {http://arxiv.org/abs/1206.4987}
  {1206.4987} \BibitemShut {NoStop}%
\bibitem [{Note1()}]{Note1}%
  \BibitemOpen
  \bibinfo {note} {Https://www.ning.com/}\BibitemShut {NoStop}%
\bibitem [{\citenamefont {Rossi}\ and\ \citenamefont
  {Ahmed}(2015)}]{rossi:2015a}%
  \BibitemOpen
  \bibfield  {author} {\bibinfo {author} {\bibfnamefont {R.~A.}\ \bibnamefont
  {Rossi}}\ and\ \bibinfo {author} {\bibfnamefont {N.~K.}\ \bibnamefont
  {Ahmed}},\ }in\ \href {http://networkrepository.com} {\emph {\bibinfo
  {booktitle} {Proceedings of the Twenty-Ninth AAAI Conference on Artificial
  Intelligence}}}\ (\bibinfo {year} {2015})\BibitemShut {NoStop}%
\bibitem [{\citenamefont {Jerome}(2013)}]{jerome:2013a}%
  \BibitemOpen
  \bibfield  {author} {\bibinfo {author} {\bibfnamefont {K.}~\bibnamefont
  {Jerome}},\ }in\ \href {http://konect.uni-koblenz.de} {\emph {\bibinfo
  {booktitle} {Proceedings Conference on World Wide Web Companion}}}\ (\bibinfo
  {year} {2013})\ pp.\ \bibinfo {pages} {1343--1350}\BibitemShut {NoStop}%
\bibitem [{\citenamefont {Leskovec}\ and\ \citenamefont
  {Krevl}(2014)}]{snapnets}%
  \BibitemOpen
  \bibfield  {author} {\bibinfo {author} {\bibfnamefont {J.}~\bibnamefont
  {Leskovec}}\ and\ \bibinfo {author} {\bibfnamefont {A.}~\bibnamefont
  {Krevl}},\ }\href@noop {} {\enquote {\bibinfo {title} {{SNAP Datasets}:
  {Stanford} large network dataset collection},}\ } (\bibinfo {year}
  {2014})\BibitemShut {NoStop}%
\bibitem [{\citenamefont {Dao}\ \emph {et~al.}(2017)\citenamefont {Dao},
  \citenamefont {Bothorel},\ and\ \citenamefont {Lenca}}]{dao:2017a}%
  \BibitemOpen
  \bibfield  {author} {\bibinfo {author} {\bibfnamefont {V.}~\bibnamefont
  {Dao}}, \bibinfo {author} {\bibfnamefont {C.}~\bibnamefont {Bothorel}}, \
  and\ \bibinfo {author} {\bibfnamefont {P.}~\bibnamefont {Lenca}},\ }in\
  \href@noop {} {\emph {\bibinfo {booktitle} {NetSci-X 2017 : 3rd International
  Winter School and Conference on Network Science}}},\ Vol.\ \bibinfo {volume}
  {Springer Proceedings in Complexity}\ (\bibinfo {year} {2017})\ pp.\ \bibinfo
  {pages} {11--19}\BibitemShut {NoStop}%
\bibitem [{\citenamefont {Erd{\H o}s}\ and\ \citenamefont
  {R{\'e}nyi}(1959)}]{erdos:1959a}%
  \BibitemOpen
  \bibfield  {author} {\bibinfo {author} {\bibfnamefont {P.}~\bibnamefont
  {Erd{\H o}s}}\ and\ \bibinfo {author} {\bibfnamefont {A.}~\bibnamefont
  {R{\'e}nyi}},\ }\href@noop {} {\bibfield  {journal} {\bibinfo  {journal}
  {Publicationes Mathematicae (Debrecen)}\ }\textbf {\bibinfo {volume} {6}},\
  \bibinfo {pages} {290} (\bibinfo {year} {1959})}\BibitemShut {NoStop}%
\bibitem [{\citenamefont {{Newman}}\ \emph {et~al.}(2001)\citenamefont
  {{Newman}}, \citenamefont {{Strogatz}},\ and\ \citenamefont
  {{Watts}}}]{newman:2001a}%
  \BibitemOpen
  \bibfield  {author} {\bibinfo {author} {\bibfnamefont {M.~E.~J.}\
  \bibnamefont {{Newman}}}, \bibinfo {author} {\bibfnamefont {S.~H.}\
  \bibnamefont {{Strogatz}}}, \ and\ \bibinfo {author} {\bibfnamefont {D.~J.}\
  \bibnamefont {{Watts}}},\ }\href {\doibase 10.1103/PhysRevE.64.026118}
  {\bibfield  {journal} {\bibinfo  {journal} {Phys. Rev. E}\ }\textbf {\bibinfo
  {volume} {64}},\ \bibinfo {eid} {026118} (\bibinfo {year} {2001})},\ \Eprint
  {http://arxiv.org/abs/cond-mat/0007235} {cond-mat/0007235} \BibitemShut
  {NoStop}%
\bibitem [{\citenamefont {{Watts}}\ and\ \citenamefont
  {Strogatz}(1998)}]{watts:1998a}%
  \BibitemOpen
  \bibfield  {author} {\bibinfo {author} {\bibfnamefont {J.~D.}\ \bibnamefont
  {{Watts}}}\ and\ \bibinfo {author} {\bibfnamefont {H.~S.}\ \bibnamefont
  {Strogatz}},\ }\href {\doibase 10.1038/30918} {\bibfield  {journal} {\bibinfo
   {journal} {Nature}\ }\textbf {\bibinfo {volume} {393}},\ \bibinfo {pages}
  {440} (\bibinfo {year} {1998})}\BibitemShut {NoStop}%
\bibitem [{\citenamefont {{Barab{\'a}si}}\ and\ \citenamefont
  {{Albert}}(1999)}]{barabasi:1999a}%
  \BibitemOpen
  \bibfield  {author} {\bibinfo {author} {\bibfnamefont {A.-L.}\ \bibnamefont
  {{Barab{\'a}si}}}\ and\ \bibinfo {author} {\bibfnamefont {R.}~\bibnamefont
  {{Albert}}},\ }\href {\doibase 10.1126/science.286.5439.509} {\bibfield
  {journal} {\bibinfo  {journal} {Science}\ }\textbf {\bibinfo {volume}
  {286}},\ \bibinfo {pages} {509} (\bibinfo {year} {1999})},\ \Eprint
  {http://arxiv.org/abs/cond-mat/9910332} {cond-mat/9910332} \BibitemShut
  {NoStop}%
\bibitem [{\citenamefont {{Klemm}}\ and\ \citenamefont
  {{Egu{\'{\i}}luz}}(2002)}]{klemm:2002a}%
  \BibitemOpen
  \bibfield  {author} {\bibinfo {author} {\bibfnamefont {K.}~\bibnamefont
  {{Klemm}}}\ and\ \bibinfo {author} {\bibfnamefont {V.~M.}\ \bibnamefont
  {{Egu{\'{\i}}luz}}},\ }\href {\doibase 10.1103/PhysRevE.65.057102} {\bibfield
   {journal} {\bibinfo  {journal} {{Phys. Rev. E}}\ }\textbf {\bibinfo {volume}
  {65}},\ \bibinfo {eid} {057102} (\bibinfo {year} {2002})},\ \Eprint
  {http://arxiv.org/abs/cond-mat/0107607} {cond-mat/0107607} \BibitemShut
  {NoStop}%
\bibitem [{\citenamefont {{Fronczak}}\ \emph {et~al.}(2003)\citenamefont
  {{Fronczak}}, \citenamefont {{Fronczak}},\ and\ \citenamefont
  {{Ho{\l}yst}}}]{fronczak:2003a}%
  \BibitemOpen
  \bibfield  {author} {\bibinfo {author} {\bibfnamefont {A.}~\bibnamefont
  {{Fronczak}}}, \bibinfo {author} {\bibfnamefont {P.}~\bibnamefont
  {{Fronczak}}}, \ and\ \bibinfo {author} {\bibfnamefont {J.~A.}\ \bibnamefont
  {{Ho{\l}yst}}},\ }\href {\doibase 10.1103/PhysRevE.68.046126} {\bibfield
  {journal} {\bibinfo  {journal} {{Phys. Rev. E}}\ }\textbf {\bibinfo {volume}
  {68}},\ \bibinfo {eid} {046126} (\bibinfo {year} {2003})},\ \Eprint
  {http://arxiv.org/abs/cond-mat/0306255} {cond-mat/0306255} \BibitemShut
  {NoStop}%
\bibitem [{\citenamefont {Newman}(2010)}]{newman:2010a}%
  \BibitemOpen
  \bibfield  {author} {\bibinfo {author} {\bibfnamefont {M.~E.~J.}\
  \bibnamefont {Newman}},\ }\enquote {\bibinfo {title} {Networks: An
  introduction},}\ \ (\bibinfo  {publisher} {Oxford University Press},\
  \bibinfo {year} {2010})\BibitemShut {NoStop}%
\bibitem [{\citenamefont {{Lancichinetti}}\ \emph {et~al.}(2008)\citenamefont
  {{Lancichinetti}}, \citenamefont {{Fortunato}},\ and\ \citenamefont
  {{Radicchi}}}]{LFR:2008}%
  \BibitemOpen
  \bibfield  {author} {\bibinfo {author} {\bibfnamefont {A.}~\bibnamefont
  {{Lancichinetti}}}, \bibinfo {author} {\bibfnamefont {S.}~\bibnamefont
  {{Fortunato}}}, \ and\ \bibinfo {author} {\bibfnamefont {F.}~\bibnamefont
  {{Radicchi}}},\ }\href {\doibase 10.1103/PhysRevE.78.046110} {\bibfield
  {journal} {\bibinfo  {journal} {{Phys. Rev. E}}\ }\textbf {\bibinfo {volume}
  {78}},\ \bibinfo {pages} {046110} (\bibinfo {year} {2008})},\ \Eprint
  {http://arxiv.org/abs/0805.4770} {0805.4770 [physics.soc-ph]} \BibitemShut
  {NoStop}%
\end{thebibliography}
\end{document}